\documentclass[longauth,letter]{aa}  

\usepackage{hyperref}
\usepackage{soul}

\usepackage{graphicx}
\usepackage{txfonts}

\usepackage{xspace,upgreek}

\def\Sunits{\ensuremath{k_{\rm B}\,{\rm baryon}^{-1}}\xspace}

\def\MJ{\ensuremath{M_{\rm Jup}}\xspace}
\def\RJ{\ensuremath{R_{\rm Jup}}\xspace}
\def\Teff{\ensuremath{T_{\rm eff}}\xspace}
\def\Lbol{\ensuremath{L_{\rm bol}}\xspace}

\usepackage{etoolbox}

\makeatletter

\patchcmd{\NAT@citex}
  {\@citea\NAT@hyper@{%
     \NAT@nmfmt{\NAT@nm}%
     \hyper@natlinkbreak{\NAT@aysep\NAT@spacechar}{\@citeb\@extra@b@citeb}%
     \NAT@date}}
  {\@citea\NAT@nmfmt{\NAT@nm}%
   \NAT@aysep\NAT@spacechar\NAT@hyper@{\NAT@date}}{}{}

\patchcmd{\NAT@citex}
  {\@citea\NAT@hyper@{%
     \NAT@nmfmt{\NAT@nm}%
     \hyper@natlinkbreak{\NAT@spacechar\NAT@@open\if*#1*\else#1\NAT@spacechar\fi}%
       {\@citeb\@extra@b@citeb}%
     \NAT@date}}
  {\@citea\NAT@nmfmt{\NAT@nm}%
   \NAT@spacechar\NAT@@open\if*#1*\else#1\NAT@spacechar\fi\NAT@hyper@{\NAT@date}}
  {}{}

\makeatother

\makeatletter
\renewcommand*\aa@pageof{, page \thepage{} of \pageref*{LastPage}}
\makeatother

\defcitealias{saumonEvolutionDwarfsColorMagnitude2008}{SM08}
\def\SMref{\citetalias{saumonEvolutionDwarfsColorMagnitude2008}\xspace}

\begin{document}

   \title{Direct Discovery of the Inner Exoplanet in the HD\,206893 System 
   }

   \subtitle{Evidence for Deuterium Burning in a Planetary-Mass Companion}

   \author{S.~Hinkley\inst{\ref{exeter}}
 \and S.~Lacour\inst{\ref{lesia},\ref{esog}}
 \and G.-D.~Marleau\inst{\ref{duisburg},\ref{tuebingen},\ref{bern},\ref{mpia}}
 \and A.-M.~Lagrange\inst{\ref{ipag},\ref{lesia}}
 \and J.~J.~Wang\inst{\ref{caltech}}
 \and J.~Kammerer\inst{\ref{stsci}}
 \and A.~Cumming\inst{\ref{mcgill}, \ref{irex}}
 \and M.~Nowak\inst{\ref{cam}}
 \and L.~Rodet\inst{\ref{cornell}}
 \and T.~Stolker\inst{\ref{leiden}}
 \and W.-O.~Balmer\inst{\ref{jhupa},\ref{stsci}}
 \and S.~Ray\inst{\ref{exeter}}
 \and M.~Bonnefoy\inst{\ref{ipag}}
 \and P.~Molli\`ere\inst{\ref{mpia}}
 \and C.~Lazzoni\inst{\ref{exeter}}
 \and G.~Kennedy\inst{\ref{warwick}}
 \and C.~Mordasini\inst{\ref{bern}}
 \and R.~Abuter\inst{\ref{esog}}
  \and S.~Aigrain \inst{\ref{oxford}}                               
 \and A.~Amorim\inst{\ref{lisboa},\ref{centra}}
 \and R.~Asensio-Torres\inst{\ref{mpia}}
 \and C.~Babusiaux\inst{\ref{ipag},\ref{lesia}}                   
 \and M.~Benisty\inst{\ref{ipag}}
 \and J.-P.~Berger\inst{\ref{ipag}}
 \and H.~Beust\inst{\ref{ipag}}                                   
 \and S.~Blunt\inst{\ref{caltech}}
 \and A.~Boccaletti\inst{\ref{lesia}}
 \and A.~Bohn\inst{\ref{leiden}}
 \and H.~Bonnet\inst{\ref{esog}}
 \and G.~Bourdarot\inst{\ref{mpe},\ref{ipag}}
 \and W.~Brandner\inst{\ref{mpia}}
 \and F.~Cantalloube\inst{\ref{lam}}
 \and P.~Caselli \inst{\ref{mpe}}
 \and B.~Charnay\inst{\ref{lesia}}
 \and G.~Chauvin\inst{\ref{ipag}}
 \and A.~Chomez\inst{\ref{lesia},\ref{ipag}}                 
 \and E.~Choquet\inst{\ref{lam}}
 \and V.~Christiaens\inst{\ref{monash}}
 \and Y.~Cl\'enet\inst{\ref{lesia}}
 \and V.~Coud\'e~du~Foresto\inst{\ref{lesia}}
 \and A.~Cridland\inst{\ref{leiden}}
 \and P.~Delorme \inst{\ref{ipag}}                           
 \and R.~Dembet\inst{\ref{lesia}}
 \and A.~Drescher\inst{\ref{mpe}}
 \and G.~Duvert\inst{\ref{ipag}}
 \and A.~Eckart\inst{\ref{cologne},\ref{bonn}}
 \and F.~Eisenhauer\inst{\ref{mpe}}
 \and H.~Feuchtgruber\inst{\ref{mpe}}
 \and F.~Galland\inst{\ref{ipag}}                         
 \and P.~Garcia\inst{\ref{centra},\ref{porto}}
 \and R.~Garcia~Lopez\inst{\ref{dublin},\ref{mpia}}
 \and T.~Gardner\inst{\ref{umich}}
 \and E.~Gendron\inst{\ref{lesia}}
 \and R.~Genzel\inst{\ref{mpe}}
 \and S.~Gillessen\inst{\ref{mpe}}
 \and J.~H.~Girard\inst{\ref{stsci}}
 \and A. Grandjean \inst{\ref{ipag}}                  
 \and X.~Haubois\inst{\ref{esoc}}
 \and G.~Hei\ss el\inst{\ref{lesia}\ref{estec}}
 \and Th.~Henning\inst{\ref{mpia}}
 \and S.~Hippler\inst{\ref{mpia}}
 \and M.~Horrobin\inst{\ref{cologne}}
 \and M.~Houll\'e\inst{\ref{lam}}
 \and Z.~Hubert\inst{\ref{ipag}}
 \and L.~Jocou\inst{\ref{ipag}}
 \and M.~Keppler\inst{\ref{mpia}}
 \and P.~Kervella\inst{\ref{lesia}}
 \and L.~Kreidberg\inst{\ref{mpia}}
 \and V.~Lapeyr\`ere\inst{\ref{lesia}}
 \and J.-B.~Le~Bouquin\inst{\ref{ipag}}
 \and P.~L\'ena\inst{\ref{lesia}}
 \and D.~Lutz\inst{\ref{mpe}}
 \and A.-L.~Maire\inst{\ref{ipag}}
 \and F.~Mang\inst{\ref{mpe}}
 \and A.~M\'erand\inst{\ref{esog}}
 \and N.~Meunier\inst{\ref{ipag}}                          
 \and J.~D.~Monnier\inst{\ref{umich}}
 \and D.~Mouillet\inst{\ref{ipag}}
 \and E.~Nasedkin\inst{\ref{mpia}}
 \and T.~Ott\inst{\ref{mpe}}
 \and G.~P.~P.~L.~Otten\inst{\ref{lam}, \ref{asiaa}}
 \and C.~Paladini\inst{\ref{esoc}}
 \and T.~Paumard\inst{\ref{lesia}}
 \and K.~Perraut\inst{\ref{ipag}}
 \and G.~Perrin\inst{\ref{lesia}} 
 \and F.~Philipot\inst{\ref{lesia}}                        
 \and O.~Pfuhl\inst{\ref{esog}}
 \and N.~Pourr\'e\inst{\ref{ipag}}
 \and L.~Pueyo\inst{\ref{stsci}}
 \and J.~Rameau\inst{\ref{ipag}}
 \and E.~Rickman\inst{\ref{esa}}
 \and P. Rubini\inst{\ref{pixyl}}                            
 \and Z.~Rustamkulov \inst{\ref{jhueps}}
 \and M.~Samland \inst{\ref{mpia}}
 \and J.~Shangguan\inst{\ref{mpe}}
 \and T.~Shimizu \inst{\ref{mpe}}
 \and D.~Sing \inst{\ref{jhupa},\ref{jhueps}}
 \and C.~Straubmeier\inst{\ref{cologne}}
 \and E.~Sturm\inst{\ref{mpe}}
 \and L.~J.~Tacconi\inst{\ref{mpe}}
 \and E.F.~van~Dishoeck\inst{\ref{leiden},\ref{mpe}}
 \and A.~Vigan\inst{\ref{lam}}
 \and F.~Vincent\inst{\ref{lesia}}
 \and K.~Ward-Duong\inst{\ref{smith}}
 \and F.~Widmann\inst{\ref{mpe}}
 \and E.~Wieprecht\inst{\ref{mpe}}
 \and E.~Wiezorrek\inst{\ref{mpe}}
 \and J.~Woillez\inst{\ref{esog}}
 \and S.~Yazici\inst{\ref{mpe}}
 \and A.~Young\inst{\ref{mpe}}
 \and N.~Zicher\inst{\ref{oxford}}       
 \and  the GRAVITY Collaboration}
\institute{ 
   University of Exeter, Physics Building, Stocker Road, Exeter EX4 4QL, United Kingdom
\label{exeter}      \and
   LESIA, Observatoire de Paris, Universit\'e PSL, CNRS, Sorbonne Universit\'e, Universit\'e Paris Cit\'e, 5 place Jules Janssen, 92195 Meudon, France
\label{lesia}      \and
   European Southern Observatory, Karl-Schwarzschild-Stra\ss e 2, 85748 Garching, Germany
\label{esog}      \and
   Fakult\"at f\"ur Physik, Universit\"at Duisburg-Essen, Lotharstraße 1, 47057 Duisburg, Germany
\label{duisburg}      \and
   Instit\"ut f\"ur Astronomie und Astrophysik, Universit\"at T\"ubingen, Auf der Morgenstelle 10, 72076 T\"ubingen, Germany
\label{tuebingen}      \and
      Center for Space and Habitability, Universit\"at Bern, Gesellschaftsstrasse 6, 3012 Bern, Switzerland
\label{bern}      \and
   Max-Planck-Institut f\"ur Astronomie, K\"onigstuhl 17, 69117 Heidelberg, Germany
\label{mpia}      \and
   Universit\'e Grenoble Alpes, CNRS, IPAG, 38000 Grenoble, France
\label{ipag}      \and
   Department of Astronomy, California Institute of Technology, Pasadena, CA 91125, USA
\label{caltech}      \and
   Space Telescope Science Institute, Baltimore, MD 21218, USA
\label{stsci}      \and
   Institute of Astronomy, University of Cambridge, Madingley Road, Cambridge CB3 0HA, United Kingdom
\label{cam}      \and
   Center for Astrophysics and Planetary Science, Department of Astronomy, Cornell University, Ithaca, NY 14853, USA
\label{cornell}      \and
   Department of Physics \& Astronomy, Johns Hopkins University, 3400 N. Charles Street, Baltimore, MD 21218, USA
\label{jhupa}      \and
   Leiden Observatory, Leiden University, P.O. Box 9513, 2300 RA Leiden, The Netherlands
\label{leiden}      \and
   Department of Physics, University of Warwick, Coventry CV4 7AL, UK
\label{warwick}     \and
   Universidade de Lisboa - Faculdade de Ci\^encias, Campo Grande, 1749-016 Lisboa, Portugal
\label{lisboa}      \and
   CENTRA - Centro de Astrof\' isica e Gravita\c c\~ao, IST, Universidade de Lisboa, 1049-001 Lisboa, Portugal
\label{centra}      \and
   Max Planck Institute for extraterrestrial Physics, Giessenbachstra\ss e~1, 85748 Garching, Germany
\label{mpe}      \and
   Aix Marseille Univ, CNRS, CNES, LAM, Marseille, France
\label{lam}      \and
   School of Physics and Astronomy, Monash University, Clayton, VIC 3800, Melbourne, Australia
\label{monash}      \and
   1. Institute of Physics, University of Cologne, Z\"ulpicher Stra\ss e 77, 50937 Cologne, Germany
\label{cologne}      \and
   Max Planck Institute for Radio Astronomy, Auf dem H\"ugel 69, 53121 Bonn, Germany
\label{bonn}      \and
   Universidade do Porto, Faculdade de Engenharia, Rua Dr. Roberto Frias, 4200-465 Porto, Portugal
\label{porto}      \and
   School of Physics, University College Dublin, Belfield, Dublin 4, Ireland
\label{dublin}      \and
   Astronomy Department, University of Michigan, Ann Arbor, MI 48109 USA
\label{umich}      \and
   European Southern Observatory, Casilla 19001, Santiago 19, Chile
\label{esoc}      \and
   European Space Agency (ESA), ESA Office, Space Telescope Science Institute, 3700 San Martin Drive, Baltimore, MD 21218, USA
\label{esa}      \and
   Department of Earth \& Planetary Sciences, Johns Hopkins University, Baltimore, MD, USA
\label{jhueps}      \and
   Department of Astronomy, Smith College, Northampton MA 01063 USA
\label{smith}    \and 
   Pixyl S.A.\ La Tronche, France
\label{pixyl}     \and 
   Department of Physics, University of Oxford, UK 
\label{oxford}     \and 
    Department of Physics and McGill Space Institute, McGill University, 3600 rue University, Montreal QC H3A 2T8, Canada 
\label{mcgill}     \and 
    Institut de Recherche sur les Exoplan\`etes (iREx), Universit\'e de Montr\'eal, C.P.\ 6128 Succ.\ Centre-ville, Montreal, QC H3C 3J7, Canada
\label{irex}        \and 
    Academia Sinica, Institute of Astronomy and Astrophysics, 11F Astronomy-Mathematics Building, NTU/AS campus, No. 1, Section 4, Roosevelt Rd., Taipei 10617, Taiwan
\label{asiaa}       \and 
    Advanced Concepts Team, European Space Agency, TEC-SF, ESTEC, Keplerlaan 1, 2201 AZ Noordwijk, The Netherlands
\label{estec}
}

   \date{Received 10 August 2022; accepted 25 October 2022}


  \abstract
   {}
   {HD\,206893 is a nearby debris disk star hosting a previously identified brown dwarf companion with a $\sim$10\,au orbital separation. Long term precise radial velocity (RV) monitoring, as well as anomalies in the system proper motion, have suggested the presence of an additional, inner companion in the system.  
   }
   {Using information from ongoing precision RV measurements with the HARPS spectrograph, as well as \textit{Gaia} host star astrometry, we have undertaken a multi-epoch search for the purported additional planet using the VLTI/GRAVITY instrument.}
   {We report a high significance detection of the companion HD\,206893c over three epochs, which shows clear evidence for Keplerian orbital motion.  Our astrometry with $\sim$50--100\,$\upmu$arcsec precision afforded by GRAVITY allows us to derive a dynamical mass of 12.7$^{+1.2}_{-1.0}$\,\MJ and an orbital separation of 3.53$^{+0.08}_{-0.06}$\,au for HD\,206893c.  Our fits to the orbits of both companions in the system use both \textit{Gaia} astrometry and RVs to also provide a precise dynamical estimate of the previously uncertain mass of the B component, and therefore allow us to derive an age of 155$\pm$15\,Myr for the system. We find that theoretical atmospheric/evolutionary models incorporating deuterium burning for HD\,206893c, parameterized by cloudy atmosphere models as well as a ``hybrid sequence'' (encompassing a transition from cloudy to cloud-free) provide a good simultaneous fit to the luminosity of both HD\,206893B and c.  
   Thus, accounting for both deuterium burning as well as clouds is crucial to understanding the luminosity evolution of HD\,206893c.
   }
   {In addition to using long-term RV information, this effort is an early example of a direct imaging discovery of a bona fide exoplanet that was guided in part with \textit{Gaia} astrometry. 
   Utilizing \textit{Gaia} astrometry is expected to be one of the primary techniques going forward to identify and characterize additional directly imaged planets.  
   In addition, HD\,206893c is an example of an object narrowly straddling the deuterium-burning limit, but unambiguously undergoing deuterium burning.
   Additional discoveries like this may therefore help to clarify the discrimination between an object that would be considered a brown dwarf or an extrasolar planet.  
   Lastly, this discovery is another example of the power of optical interferometry to directly detect and characterize extrasolar planets where they form at ice-line orbital separations of 2--3\,au.
   }


   \maketitle
%


\begin{figure*}[ht]
		\centering
		\includegraphics[width=0.53\linewidth]{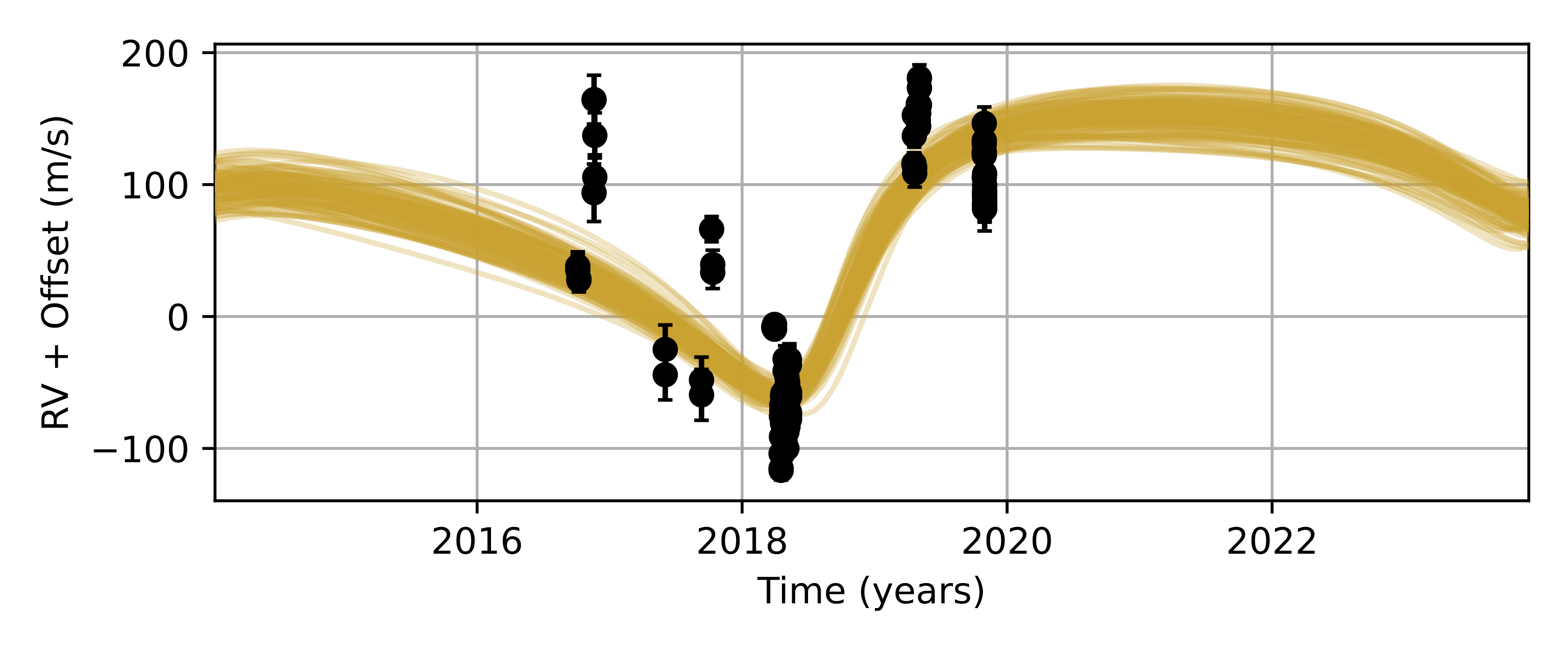} \hspace{0.1in}
		\includegraphics[width=0.44\linewidth]{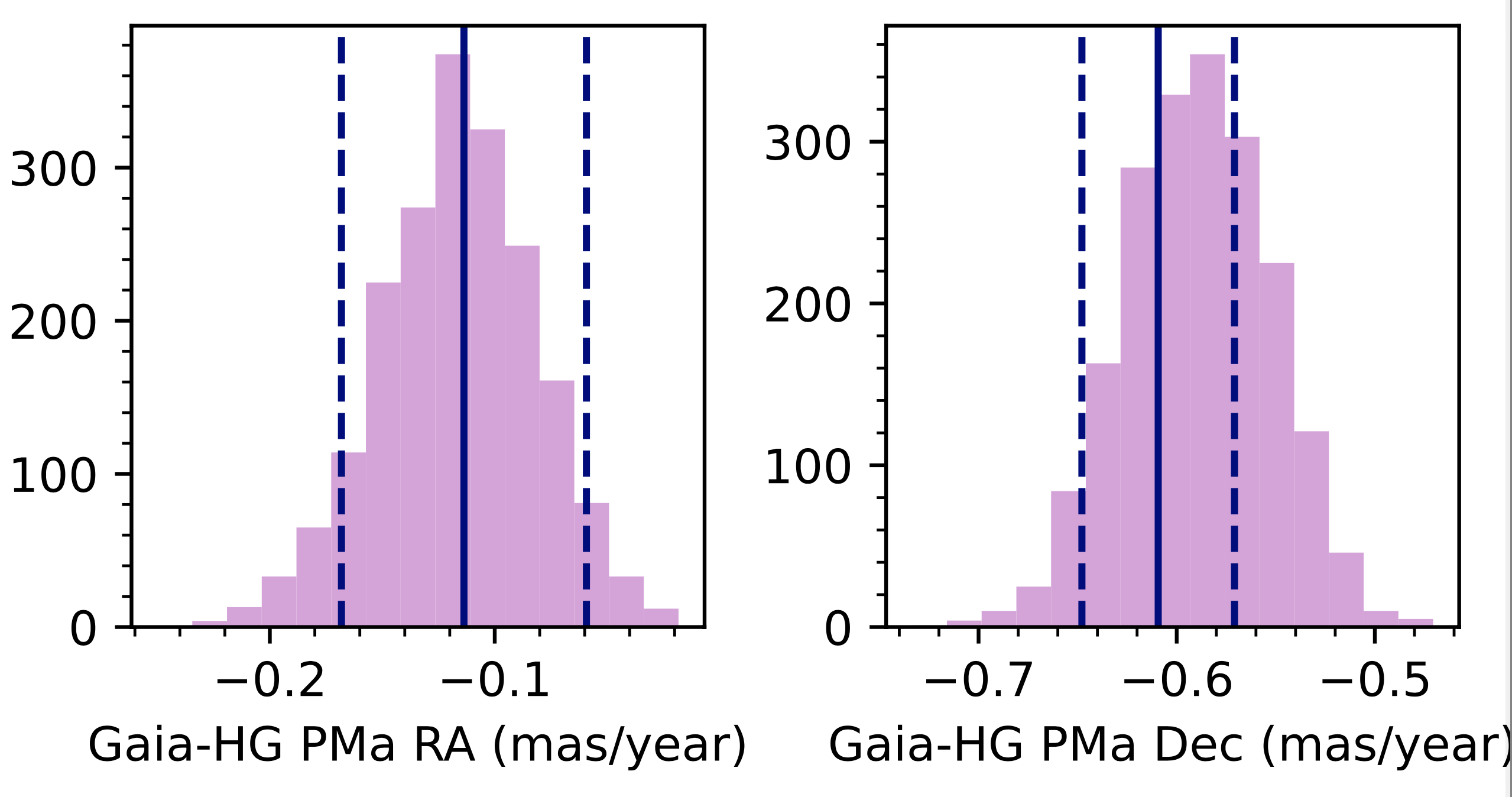}
		\caption{\textit{Left:} The HARPS radial velocities for HD\,206893 \citep[e.g.][]{glb19} along with samples from the posterior for the model fit to the radial velocities and proper motion anomalies. \textit{Right:} The posterior distributions for our best fit model of the predicted proper motion anomaly (purple histograms) calculated as the difference between the \textit{Gaia} EDR3 proper motion and the long-term proper motion as calculated by comparing \textit{Gaia} with \textit{Hipparcos} (labelled ``HG'' in the figure) and displayed individually in terms of Right Ascension and Declination.  The vertical blue lines show the actual values, with 1$\sigma$ uncertainties represented by the dashed lines. }
		\label{fig:rv_and_PMA}
	\end{figure*}

\section{Introduction} 
\vspace{-0.2cm}
What distinguishes an extrasolar giant planet (EGP) from a brown dwarf (BD) is still a matter of debate.  The current IAU definition of a planet is an object that is below the mass required for thermonuclear fusion of deuterium, which we currently believe occurs at 13 Jupiter masses. 
Identifying objects near this mass limit will clarify how distinct the boundary between massive planets and brown dwarfs really is \citep[e.g.,][]{mm12, bdl13}. However, what qualifies an object to be an exoplanet may be more related to its formation mechanism rather than solely its mass \citep[e.g.,][]{cbs07}. 
Hybrid exoplanetary systems containing both a brown dwarf as well as an exoplanet, which presumably formed at the same time from the same protoplanetary disk, are therefore ideal systems to study different possible formation pathways using diagnostics like atmospheric atomic ratios \citep[e.g.,][]{omb11, mak14, mvm16, mmb22} or initial entropy \citep[e.g.,][]{sb12,mc14}.  Finding and characterizing planetary mass companions near this deuterium burning limit \citep[e.g.,][]{bls13,bcm14,hpf13}, or systems containing both a brown dwarf and an exoplanet, will allow us to clarify how we draw a distinction between these two classes of objects \citep[e.g.,][]{mm12}.

HD\,206893 is a nearby \citep[40.707$\pm$0.067\,pc,][]{gbv18} F5V star with roughly solar metallicity ranging from $\mathrm{[Fe/H]}=-0.07$ to $-0.05$ \citep{hna07,grb16} to $+0.01$ \citep{n17}, and a highly uncertain age due to the lack of clear membership to any young kinematic moving groups. Most age estimates of this star have been in the range of 100--300\,Myr \citep{zs04dusty, dsb17, mhc17,kls21}, but some works have allowed ages as young as 3 or 50\,Myr \citep{kls21,wpf21}, while another has quoted an age as high as 2.1\,Gyr \citep[e.g.,][]{dh15}.  A substellar companion, HD\,206893B was first identified in \citet{mhc17} showing a projected orbital separation of $\sim$10\,au using the VLT SPHERE instrument.  The somewhat uncertain system age, as well as an unconstrained astrometric orbit, allowed only loose constraints on the mass of HD\,206893B, initially placing it in the range of 24--73\,\MJ \citep{mhc17}, but later refined to be as low as $\sim$5\,\MJ \citep{kls21} and up to 30--40\,\MJ \citep[][]{dsb17,kls21,wpf21}.  
This object possesses extraordinarily red infrared colors, and has been characterized in numerous subsequent works \citep{dsb17,sqt20,mgl21,wpf21,kls21}.  

Evidence has been emerging for an additional, inner companion in the system. Using precise radial velocity (RV) monitoring with the \textit{HARPS} spectrograph over 1.6\,yr, \citet{glb19} revealed a significant RV drift that could not be solely due to HD\,206893B, and hypothesized the presence of a $\sim$15\,\MJ companion.  Similarly, \citet{kls21} showed that the anomaly in the \textit{Gaia} EDR3 proper motion cannot be explained by the B companion alone. They constrained the parameter space spanned by the mass and on-sky position of an additional, hypothesized 8--15\,\MJ companion. 

In this paper, using optical interferometry over several epochs, we present the discovery  of HD\,206893c, a $\sim$12-13\,\MJ exoplanet orbiting the star at 3.5\,au.  
Our study marks the first \textit{direct} detection of HD\,206893c, providing highly precise astrometry ($\sim$50--100\,$\upmu$arcsec) of the HD\,206893c orbit, as well as medium resolution ($R\sim500$) spectroscopy in the $K$-band.
In addition to obtaining a precise dynamical mass for the HD\,206893c exoplanet, our study also provides precise astrometry of the orbit of HD\,206893B. This increased precision on the HD\,206893B orbit, combined with its very well constrained luminosity, allows us to derive a well constrained system age of 155$\pm$15\,Myr. 
This discovery of HD\,206893c establishes HD\,206893 as a hybrid planetary system hosting both an exoplanet and a brown dwarf, and therefore presents itself as a valuable laboratory for studying possible formation pathways of EGPs and BDs.  


\begin{figure*}[ht]
		\centering
	\vspace{-1.00in}	\includegraphics[width=0.9\linewidth]{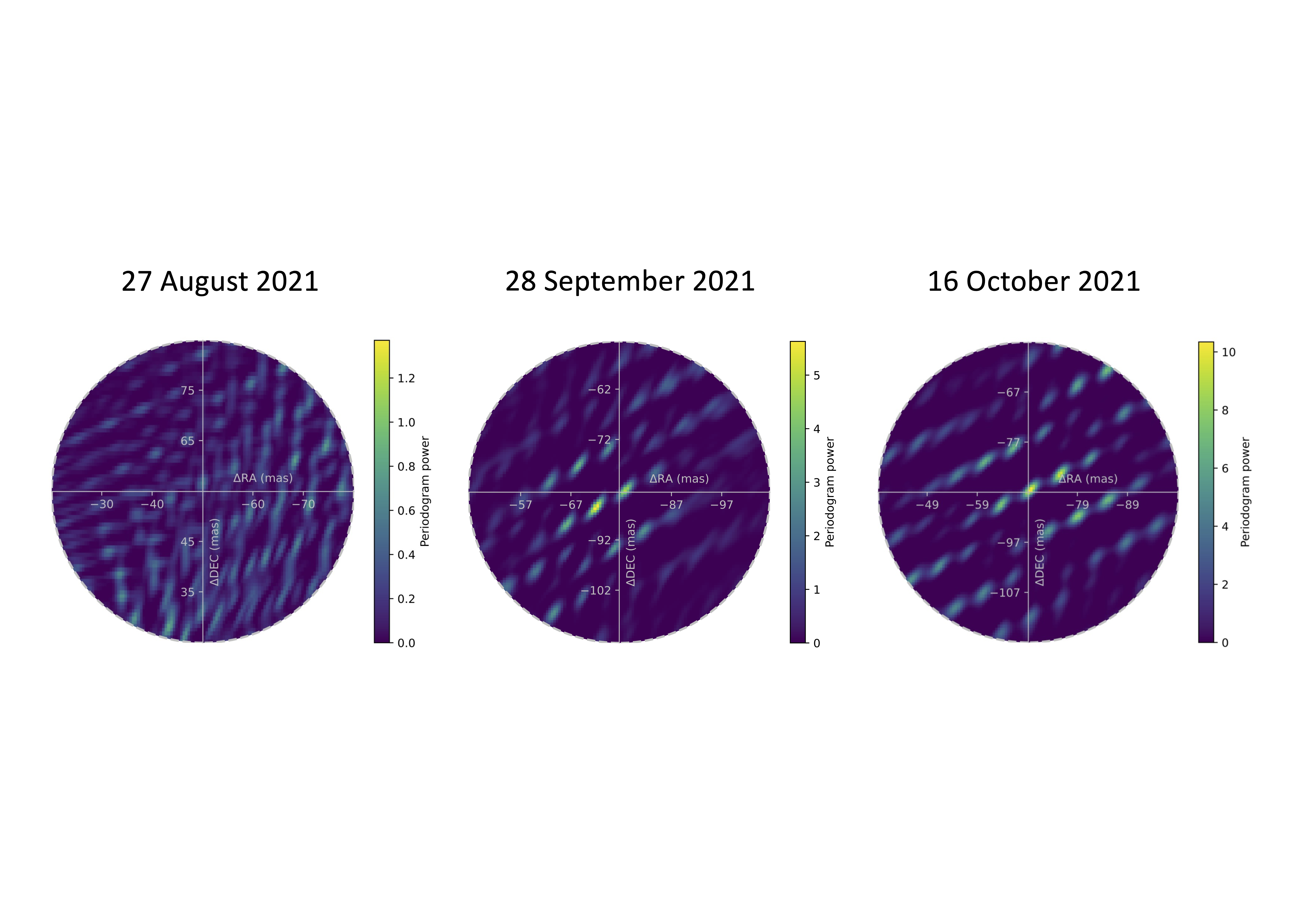}
        \vspace{-1.00in}
		\caption{Periodogram power maps, similar to the one presented in \citet{nll20}, obtained for three different observations of HD\,206893c: one resulting in a non-detection during the search for the planet (in August 2021), and two clear detections in September and October 2021.}\label{fig:detection}
\end{figure*}

\vspace{-0.3cm}
\section{Observations}

\subsection{Precision radial velocities}
As described in \citet{glb19}, HD\,206893 has been monitored for several years since 2016 with the High Accuracy Radial velocity Planet Searcher (HARPS, La Silla, ESO) as part of the Young Nearby Stars survey \citep[YNS;][]{lmc13}. The Spectroscopic data via Analysis of the Fourier Interspectrum RVs (SAFIR) software \citep[][]{glu05} was used to compute the RV signal. Figure~\ref{fig:rv_and_PMA} (\textit{left panel}) shows the processed historical RV datasets taken from \citet{glb19} that were used in this study to predict the semi-major axis of HD\,206893c. The residuals between the RV data points and the posterior curves shown in the figure are likely due to stellar pulsations.  As shown in Table~\ref{table:detailed_orbit}, there is an additional $\sigma_{RV}=40$\,m\,s$^{-1}$ amplitude best-fit stellar jitter term not currently represented in the uncertainties on the individual RV points shown in Figure~\ref{fig:rv_and_PMA}. However, as discussed in Section~\ref{sec:orbit_fits}, such residuals in the RV data will not have a significant impact on the final derived physical parameters for HD\,206893c.  

\subsection{GRAVITY observations \& data processing}\label{gravity_obs}
All the GRAVITY observations presented in this paper were acquired using the dual-field/on-axis mode, with medium spectral resolution ($R=500$). We closely followed the observing strategy outlined in \citet{gln19}, where the science fiber is alternately centered on the star to obtain the phase reference, and on the position of HD\,206893c. These observations were reduced using the procedure described in \citet{gravitycollaborationPeeringFormationHistory2020} and \citet{nll20}. The coherent flux is extracted for the on-star and on-planet exposures using the GRAVITY pipeline \citep{lkl14}. The on-planet data are then phase-referenced to the on-star observations, and the data are fitted with a model which includes both a planet component and a low-order polynomial (for speckle deconvolution). The astrometry is obtained by varying the astrometry of the planet component, and looking for the minimum $\chi^2$ or, equivalently, by maximising the periodogram power. More details are provided in Section~\ref{sec:significance} as well as Appendix~B of \citet{nll20}.

\begin{table}[]
    \centering
    \tiny
    \begin{tabular}{lccc}
         \hline
         \multicolumn{4}{c}{\textbf{Date}} \\
         Start/End UT time  & \multicolumn{1}{c}{Airmass} & $\tau_0$ & Seeing  \\  \hline
          Target & $\Delta$RA/$\Delta$DEC &  \multicolumn{2}{c}{NEXP/NDIT/DIT}  \\
         \hline
         \hline
        \multicolumn{4}{c}{\textbf{2021-08-27}}    \\
        01:14:57          /        03:41:51                &              \multicolumn{1}{c}{1.03--1.35} & 3.6--7.0\,ms   & 0.5--0.9\,$^{\prime\prime}$    \\  \hline
          HD\,206893\,A  & 0/0\,mas   & \multicolumn{2}{c}{6/64/1\,s}  \\
           HD\,206893B      &  17/201\,mas &  \multicolumn{2}{c}{4/16/30\,s }  \\
           HD\,206893\,(no det)      &  $-20$/72\,mas  &  \multicolumn{2}{c}{4/32/10\,s }  \\
           HD\,206893\,(no det)      & $-50$/54\,mas   &  \multicolumn{2}{c}{4/32/10\,s  }   \\
           HD\,206893\,(no det)      &  $-70$/30\,mas   &  \multicolumn{2}{c}{4/32/10\,s }  \\
        \hline \hline
        \multicolumn{4}{c}{\textbf{2021-08-28}}\\   
        01:55:39          /        03:55:29                &              \multicolumn{1}{c}{1.02--1.18} & 4.8--7.4\,ms   & 0.6--0.9\,$^{\prime\prime}$    \\  \hline
         HD\,206893\,A      & 0/0\,mas &  \multicolumn{2}{c}{6/64/1\,s}    \\
         HD\,206893\,(no det)      & $-72$/6\,mas&  \multicolumn{2}{c}{5/32/10\,s}       \\
        HD\,206893\,(no det)      &  $-70$/$-20$\,mas &  \multicolumn{2}{c}{5/32/10\,s}     \\
        HD\,206893c                &  $-55$/$-50$\,mas&  \multicolumn{2}{c}{5/32/10\,s}      \\
        \hline\hline
        \multicolumn{4}{c}{\textbf{2021-09-28}}\\  
        01:13:14          /        03:53:59                &              \multicolumn{1}{c}{1.00--1.15} & 3.9--3.4\,ms   & 0.8--1.2\,$^{\prime\prime}$    \\  \hline
         HD\,206893\,A      &  0/0\,mas &  \multicolumn{2}{c}{10/64/1\,s}    \\
         HD\,206893c     & $-76$/$-82$\,mas&  \multicolumn{2}{c}{11/32/10\,s}      \\
         HD\,206893\,(no det)      & $-65$/$-30$\,mas&  \multicolumn{2}{c}{8/12/30\,s}      \\
        \hline\hline
        \multicolumn{4}{c}{\textbf{2021-10-16}}\\
             23:41:45         /        03:07:57               &              \multicolumn{1}{c}{1.00--1.22} & 1.6--3.3\,ms   & 0.4--0.6\,$^{\prime\prime}$    \\  \hline
         HD\,206893\,A      & 0/0\,mas&  \multicolumn{2}{c}{12/64/1\,s   }     \\
         HD\,206893c     & $-76$/$-82$\,mas&  \multicolumn{2}{c}{27/32/10\,s}       \\
         \hline
    \end{tabular}
    \caption{Observing log of the four nights. $\tau_0$ denotes the atmospheric coherence time. The values $\Delta$RA/$\Delta$DEC are the placement of the science fiber relative to the fringe tracking fiber (which is always on the central star). NEXP, NDIT, and DIT denote the number of exposures, the number of detector integrations per exposure, and the detector integration time, respectively.  }
    \label{tab:obs}
\end{table}

Following the work from \citet{kls21}, we used the previous GRAVITY astrometry of HD\,206893B in combination with the new HARPS observations \citep{glb19,glb19corr} and \textit{Gaia} DR3 astrometry \citep{gpd16,gbv21} to predict a position for HD\,206893c in August 2021. We use the proper motion anomaly calculated as the difference between the \textit{Gaia} EDR3 proper motion and the long-term proper motion as calculated by comparing \textit{Gaia} with \textit{Hipparcos} proper motions (labelled ``HG'' in Figure~\ref{fig:rv_and_PMA}).  The previous GRAVITY astrometry measurements were particularly useful for constraining the orbit of HD\,206893B, and subsequently remove its signal from the RV data and \textit{Gaia} astrometry.  The RV data allowed us to constrain the period, and thus the semi-major axis of HD\,206893c, while the \textit{Gaia} astrometry provided constraints on its position angle.  

During the two nights between 27~and 28~August, we used the GRAVITY instrument \citep{gravitycollaborationFirstLightGRAVITY2017} to try to detect the exoplanet. GRAVITY only probes a small circular region of $\sim$50\,mas radius around the position of the single mode fiber.  We searched for the companion over a region on the order of  10\,000\,mas$^2$ by offsetting the science fiber of the instrument. At the same time the fringe tracker observed the star to correct atmospheric turbulence \citep{lacourGRAVITYFringeTracker2019}. This allowed an integration time of several tens of seconds. The log of the observations, as well as the position of the single mode fibers, are summarised in Table~\ref{tab:obs}. During the second night, on the last pointing, a weak signal was detected.

The exoplanet position was confirmed one month later, in September. A clear detection was obtained during runs ID 1104.C-0651, part of the ExoGRAVITY large program \citep{lacourExoGRAVITYProjectUsing2020}, and the confirmation during a dedicated run (ID 105.20T0.001). In October, a final observation was carried out, as part of the ExoGRAVITY large program, with the goal of obtaining a good calibration of the spectrum. Figure~\ref{fig:detection} provides examples of the periodogram power maps obtained from our observations, both for a non-detection (fiber mispointed during the search), and for a detection of the planet.

The $K$-band contrast relative to the host star for HD\,206893c is approximately 8.2$\times$10$^{-5}$, compared to 1.6$\times$10$^{-4}$ for HD\,206893B. The reliability of this contrast ratio is ensured since the host star is observed simultaneously using the GRAVITY fringe tracker, so the ratio of coherent energy is constantly monitored, and is the result of a simultaneous measurement. The main systematic uncertainty that can arise in this measurement is due to the response of the injection of the signal into the fiber, defined by the ``lobe'' of the interferometer, which can be computed analytically as described in Appendix A of \citet{wvl21}. Previous calibrations reveal that these systematics are small.

The astrometry obtained at the three epochs is given in Table~\ref{tab:astrometry}. In addition to HD\,206893c, we have recomputed the astrometry of HD\,206893B from \citet{kls21}, and added an additional position obtained in August 2021.  With respect to the theoretical performance of GRAVITY ($10\,\upmu$as, \citealp{lacourReachingMicroarcsecondAstrometry2014}), the astrometry accuracy is still a factor 5 to 20 worse. The difference during the night of 27~August can be explained by the mispointing of the science fiber while searching for the exoplanet. The error bars still do not decrease considerably during the other runs because of the limitation due to systematics, particularly the optical aberrations within the fiber injection optics \citep{gravitycollaborationImprovedGRAVITYAstrometric2021}.

\begin{table}[]
    \centering
    \begin{tabular}{cccccc}
         \hline
      {Epoch} & {$\Delta$RA} & {$\sigma_{\Delta\mathrm{RA}}$} & {$\Delta$Dec} & {$\sigma_{\Delta\mathrm{Dec}}$ } & {$\rho$} \\
      {[MJD]} & {[mas]} & {[mas]} & {[mas]} & {[mas]} &  \\
\hline\hline
\multicolumn{6}{c}{\textbf{HD\,206893B} }\\ 
\hline
58681.40$^\star$  & 130.73    & 0.06          &  198.12    & 0.06     & $-0.77$         \\
58708.16$^\star$  & 127.06  & 0.08          &  199.24    & 0.15      & $-0.93$           \\
59453.09   & 20.08   & 0.08          &  205.80    & 0.06      & $-0.87$          \\
\hline\hline
\multicolumn{6}{c}{\textbf{HD\,206893c}} \\ 
\hline
59454.12    & $-76.56$   & 0.14           & $-82.74$    & 0.09      &  $-0.88$             \\
59485.11     & $-72.06$  & 0.06           & $-85.36$     & 0.11      & $-0.83$              \\
59504.06    & $-69.27$  & 0.06           & $-86.79$    & 0.07       & $-0.70$  \\   
\hline
    \end{tabular}
    \caption{GRAVITY astrometry of HD\,206893B and HD\,206893c. The astrometry with~$\star$ are data published in \citet{kls21} but reduced with the latest version of the GRAVITY pipeline. The co-variance matrix 
    has
    $\sigma_{\Delta\mathrm{RA}}^2$ and $\sigma_{\Delta\mathrm{Dec}}^2$ on the diagonal, and $\rho\times\sigma_{\Delta\mathrm{RA}}\times\sigma_{\Delta\mathrm{Dec}}$ on the off-diagonal, where $\rho$ is the correlation coefficient.
    }
    \label{tab:astrometry}
\end{table}

\subsection{Statistical significance of the detection}\label{sec:significance}
In Fig.~\ref{fig:detection} we show an example of the periodogram power maps for three observations of HD\,206893c: one in August 2021, obtained when the fiber was mispointed during the search, and two obtained in September and October 2021, when the fiber was properly centered on the planet. Each of the three maps represents the field of view of the GRAVITY science fiber. A clear peak with some sidelobes in the periodogram power (not visible in the August 2021 observation) are characteristic of a successful detection, and give the position of the planet.

\section{Results}
\subsection{Orbital fits}\label{sec:orbit_fits}
To assess the orbital parameters for HD\,206893B and c, we employ the method presented in \citet{lwr21} to perform Bayesian parameter estimation with the \texttt{orbitize!} software package \citep{bwa20} for jointly fitting the orbits of multiple substellar companions using relative astrometry from GRAVITY, the stellar RVs described in the previous section, and Hipparcos and Gaia absolute stellar astrometry from the EDR3 edition of the HGCA \citep{b21, kat22}. This method accounts for shifts in the system barycenter due to multiple planets in the system, but does not account for planet-planet interactions, which are negligible on these timescales. The covariances of the uncertainties in RA/Dec due to $u$-$v$ coverage are also automatically handled by \texttt{orbitize!}. To find the best fit orbits for the HD\,206893B and c objects, the parallel-tempered affine-invariant sampler \citep{foreman-mackeyEmceeMCMCHammer2013,vousdenDynamicTemperatureSelection2016} was run using 20 temperatures, 1000 walkers per temperature, and 60,000 steps per walker. The walker chains were visually inspected for convergence, and the last 10,000 steps from each walker at the lowest temperature was used to form the posterior distributions for each parameter. 

\begin{figure*}[ht]%
\centering
\includegraphics[width=1.0\textwidth]{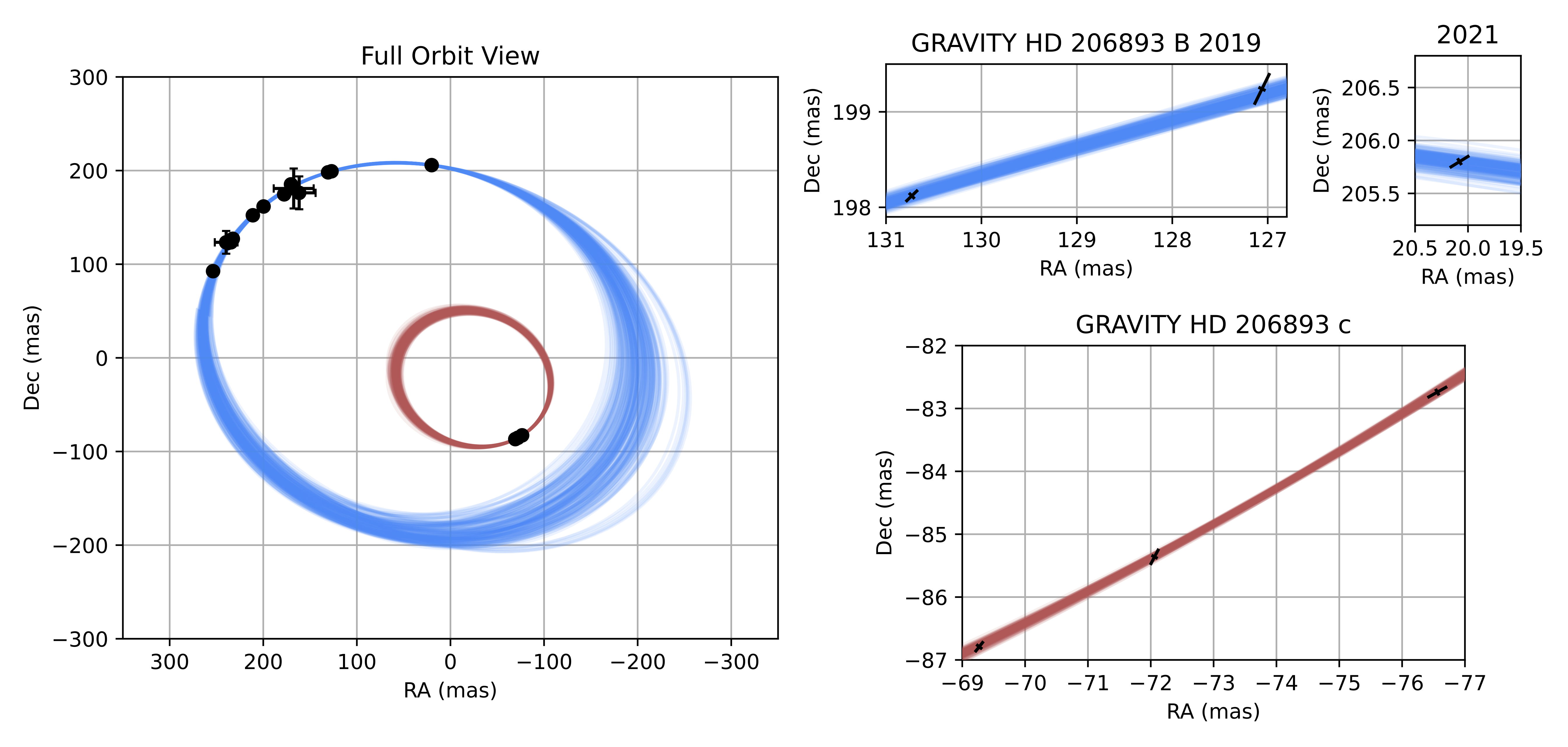}
\caption{A diagram showing the class of orbits consistent with the measured astrometry for HD\,206893B (blue orbits), and HD\,206893c (red orbits).  The HD\,206893B astrometry comes from astrometry obtained in \citet{mhc17, dsb17,glb19, sqt20, wpf21}, as well as the recent GRAVITY detections.  The right panels are magnifications (50--100x) of portions of the orbits highlighting the $\sim$50-100\,microarcsecond astrometric precision.  
}
\label{orbits}
\end{figure*}

The results of our MCMC orbital fits are summarized in Table~\ref{tab:params} and a depiction of the orbits for both HD\,206893B and c are shown in Figure~\ref{orbits}. The best estimate of the mass and semimajor axis of HD\,206893c is 12.7$^{+1.2}_{-1.0}$\,\MJ and 3.53$^{+0.08}_{-0.06}$\,au ($\sim$5.7 yr orbital period) for a dynamical fit that incorporates the precise RV monitoring from HARPS. Excluding the RV information from the fit results in values of 11.5$^{+2.4}_{-2.2}$\,\MJ and 3.68$^{+0.12}_{-0.09}$\,au. These values are largely consistent with, but also significantly more precise than, the previous estimations of these parameters for the hypothesized HD\,206983c companion of 8-15\,\MJ and $\lesssim$5.6\,au \citep{glb19,kls21}. 
At the same time, our fits reveal a precise and significantly non-zero eccentricity $e=0.41^{+0.03}_{-0.03}$ for HD\,206893c ($e=0.36^{+0.05}_{-0.06}$ when excluding the RV measurements).
For completeness we also summarize the results of our orbital fit for HD\,206893B. We find a mass and semi-major axis of $M_B=28.0^{+2.2}_{-2.1}$\,\MJ and $a_B=9.6^{+0.4}_{-0.3}$\,au corresponding to a $\sim$26-yr orbital period (26.2$^{+3.7}_{-3.6}$\,\MJ and 9.7$^{+0.8}_{-0.4}$\,au when excluding the RV measurements). More details on the orbit fitting, including the orbital alignment parameters (e.g., $i$, $\omega$, $\Omega$) can be found in Appendix~\ref{sec:orbit_fit_detail}.
As for HD\,206893c, this mass determination for HD\,206893B is consistent with, but also significantly more precise than, previous estimates \citep[e.g.][]{glb19,mgl21,wpf21,kls21} that quoted a mass of 5--30\,\MJ, and clearly places HD\,206893B in the low-mass brown dwarf regime.

\subsection[GRAVITY spectroscopy of HD\,206893c: Model grid fits to derive atmospheric parameters]{GRAVITY spectroscopy of HD\,206893c: model grid fits to derive atmospheric parameters}\label{sec:spectra_modelling}

\begin{figure*}[ht]
\centering
\includegraphics[width=15cm]{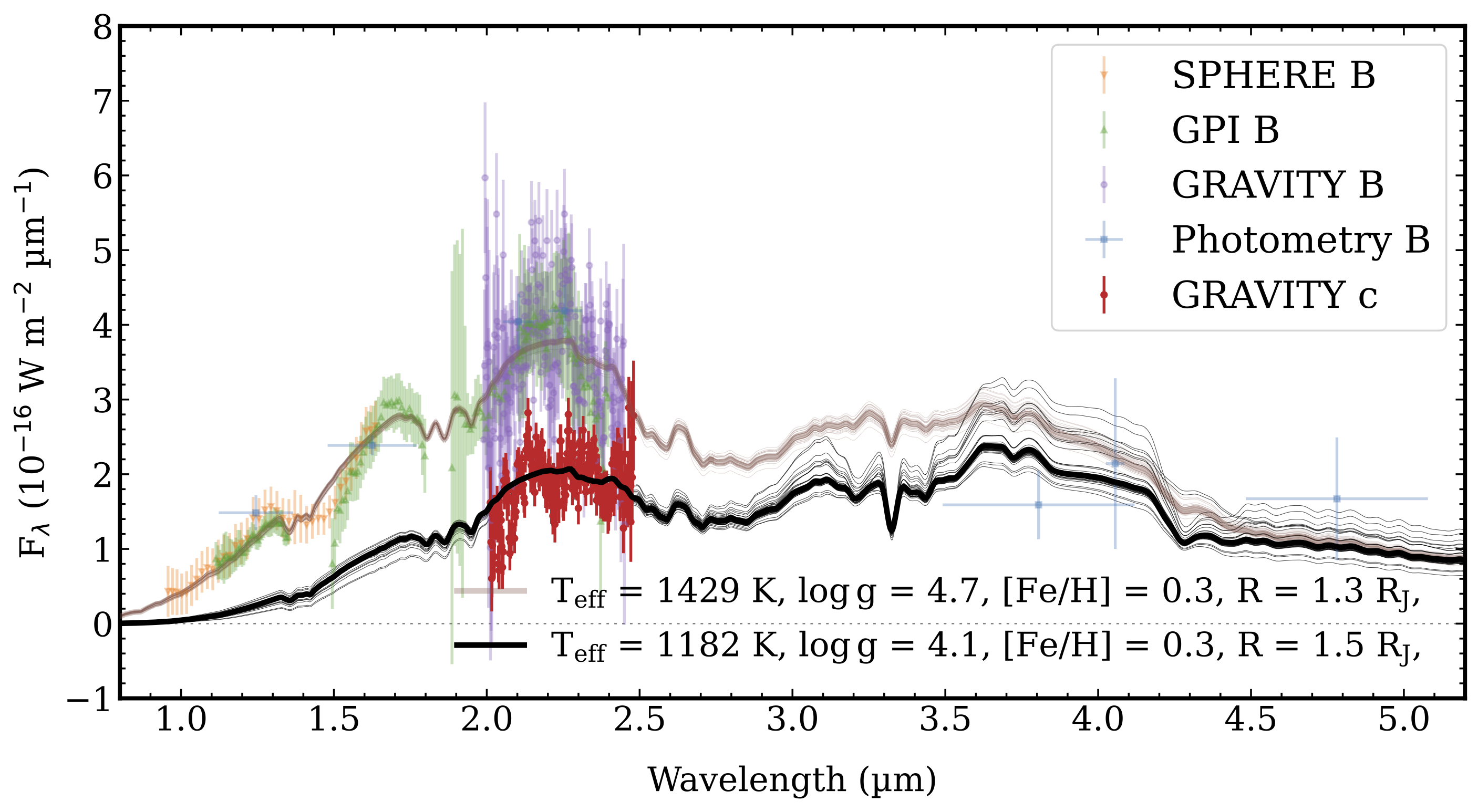}
\caption{Combined $R\sim500$ GRAVITY $K$-band spectra of HD\,206893c and B (red and light purple, respectively) together with the best fit DRIFT-PHOENIX models (black and light brown). Shown in light orange, green, and blue is archival spectrophotometry of B that was also included in the fit. For the model spectra, several samples drawn from the posterior distribution are shown with thin lines.}
\label{Jens_model}
\end{figure*}

Fig.~\ref{Jens_model} shows our collected spectrophotometry for both HD\,206893B and c.  We estimate the primary atmospheric parameters of HD\,206893c (i.e., effective temperature, surface gravity, metallicity)
with the methodology outlined in \citet{kls21}. We fit the GRAVITY spectroscopy shown in Fig.~\ref{Jens_model} with a grid of DRIFT-PHOENIX models \citep{hdw08} using the \texttt{species} toolkit \citep{sqt20}. 
We have chosen these models to fit the spectroscopy of HD\,206893c for simplicity, as well as to be consistent with those fits carried out in \citet{kls21} for the HD\,206893B companion.
Moreover, among the three different model grids used in \citet{kls21}, the DRIFT-PHOENIX grid is the only one that yields a good fit to the extremely red colors of HD\,206893B without requiring an additional source of reddening. Even if the DRIFT-PHOENIX grid does not fully capture the full physical picture of the atmosphere, the models empirically fit well, and thus are effective for measuring the bolometric luminosity.  We defer a more in-depth characterization of the atmosphere of HD\,206893c, involving a larger class of models, to a future work.

The best fit parameters for HD\,206893B and c are shown in Table~\ref{tab:params}, and the best fit DRIFT-PHOENIX spectra together with our GRAVITY spectra of both companions as well as other spectrophotometry of B from the literature are shown in Fig.~\ref{Jens_model}.
We highlight the bolometric luminosities $\log (L/L_\odot) = -4.23\pm0.01$~dex and $-4.42^{+0.02}_{-0.01}$ calculated from the best-fit models for HD\,206893B and c, respectively. We also find a best-fit metallicity $\mathrm{[Fe/H]}=0.27^{+0.02}_{-0.05}$ and $0.28^{+0.02}_{-0.04}$ for B and c, respectively, which is notably different from the nearly solar metallicity of the host star ($\textrm{[Fe/H]}=0.04\pm0.02$; \citealp{kls21}). However, for the purposes of this paper, the metallicity of the atmospheric model has a negligible effect on the final calculated bolometric luminosity, \Lbol. 
Specifically,
when the fit is restricted to solar-metallicity models ($\textrm{[Fe/H]}=0.0$), the bolometric luminosity of the best-fit model is negligibly smaller (by only 0.02~dex, or 5\,\%).
A similar difference is seen for HD\,206893B, in which the luminosity using solar-metallicity models would be $\log(L/L_\odot)= -4.20$~dex, only 0.03~dex higher than in Table~\ref{tab:params}.

Our bolometric luminosity for HD\,206893c is calculated by integrating a model that is fit only to the GRAVITY $K$-band spectrum at $\sim$2\,$\upmu$m since there are no photometric measurements at other wavelengths.To test the reliability of this \Lbol estimate,
we have calculated the bolometric luminosity for HD\,206893B
from a fit only to its $K$-band spectrum from GRAVITY, ignoring the other available data.
%
This yields $\log(L/L_\odot) = -4.26\pm0.01$, only $\lesssim$7\,\%\ smaller than the $-4.23\pm0.01$ value from the best fit
to all of the spectrophotometric data at 1--5\,$\upmu$m. Under the assumption that the spectral shapes of the two objects are similar, this suggests that our bolometric luminosity determination for HD\,206893c is robust.

The derived bolometric luminosity depends to some extent on the chosen atmospheric model. Therefore, there may be additional systematic uncertainties due to these differences between models.  
%
%
To estimate these systematics, we fit an ensemble of five additional theoretical atmosphere models (i.e., AMES-Dusty, BT-Settl, Exo-REM, petitCODE, as well as a simple blackbody curve) to the $K$-band GRAVITY spectrum of HD\,206893c.  As expected, the quality of each fit varied from model to model, and each fit returned a slightly different value for the bolometric luminosity. The rough overall standard deviation in $\log(L/L_\odot)$ is 0.20\,dex. Thus for the analysis in this paper, going forward we adopt a value of $\log(L/L_\odot) = -4.42\pm0.20$ for HD\,206893c, which should account for systematic uncertainties between models. Some of the returned radii are clearly lower than expected (but not pathologically so) but these deviations are accounted for in our presented uncertainty. We discuss the physical implications of this expanded uncertainty more in Section~\ref{sec:disc}.  


For the atmospheric model fitting, letting the mass of c be entirely free leads to a clearly incorrect estimation of the mass, with the best-fit surface gravity $\log g$ and $R$ implying $M_c{\sim}100$\,\MJ.
Therefore, we use the dynamical mass (12.7$^{+1.2}_{-1.0}$\,\MJ; Table~\ref{tab:params}) as a prior,
and obtain
9.9$^{+1.5}_{-1.7}$\,\MJ,
roughly 20\,\% smaller than the prior.
Nevertheless, the dynamical mass should be treated as much more reliable.
For the other 
atmospheric parameters ($R$, \Teff, $\log g$), we find a similar level of variation of 10--20\,\% in the values across the models. Nonetheless, the DRIFT-PHOENIX models are the most consistent with the previous model fitting in \citep{kls21}, and so we report their values in Table~\ref{tab:params}.

\subsection[Constraints on the system age and on the cloudiness of HD\,206893c]{Constraints on the system age and on the cloudiness of HD\,206893c}
 \label{sec:ageatmo}

The dynamical mass determination of HD\,206893B, precise to 10\%, can be combined with the object's bolometric luminosity to derive an age for the system by using cooling tracks. At constant age, the luminosity of hot-start gas giants scales roughly as $\Lbol\sim M^2$ at intermediate luminosities \citep{ab06,mc14}, so that the mass ratio $M_B/M_c\approx2$ would naively imply a luminosity difference of $\log(L_B/L_c)\approx0.6$~dex. However, the luminosities of HD\,206893B and c differ by only 0.2~dex.

Adjusting the initial entropy \citep{sb12, mc14} of either object
does not
bring their current luminosities closer. Indeed, an elevated initial entropy will not increase the luminosity of c at its present age because hot starts already assume an initial cooling timescale much shorter than the current age
\citep{mc14}. Similarly, no reasonable lower initial entropy for B could sufficiently delay the beginning of deuterium burning 
so that it would be observed in the rising part of a ``deuterium flash'' (see Fig.~8 of \citealt{mc14} or Fig.~13 of \citealt{bcm14}).
Given that the mass of HD\,206893c is close to the deuterium-burning limit $M\approx11.5$--$14.5~\MJ$ \citep{spiegel2011,mm12}, nuclear reactions are a likely candidate to bring the luminosity of HD\,206893c closer to that of B, and so are clouds, which may play a role in the L/T transition that low-mass objects experience \citep{Dupuy12, lda16}. Coincidentally, both can occur at similar effective temperatures \citep{saumonEvolutionDwarfsColorMagnitude2008}, depending on the mass of the object.

Therefore, we turn to the models of \citet[][hereafter \SMref]
{saumonEvolutionDwarfsColorMagnitude2008}. The \SMref models provide luminosities and magnitudes for objects cooling with a cloudy or a clear (cloud-free) atmosphere at all ages, which they show is very similar to respectively COND03 \citep{baraffeEvolutionaryModelsCool2003} or DUSTY00 \citep{chabrierEvolutionaryModelsVery2000}.
The cloudiness of the atmosphere influences the cooling rate and thus the luminosity and radius evolution of low-mass objects.
The advantage of the \SMref models is that they also provide a ``hybrid sequence'' using a transition from cloudy at high effective temperature $\Teff{\geqslant}1400$~K to cloud-free at lower $\Teff{\leqslant}1200$~K. This is a simplified model of the L/T transition but has the potential of delaying the cooling differently for HD\,206893B and~c. Despite the numerous improvements in opacities since \citeyear{saumonEvolutionDwarfsColorMagnitude2008}, the recent cloud-free {Sonora} models \citep{marleySonora21} have very similar isochrones to \SMref or COND03. Given this, and since cloudy {Sonora} models are not yet available, the \SMref models are a good modeling choice, and continue to be applied to modelling substellar objects (e.g., VHS\,J1256--1257AB\,b; \citealp{dupuy22_1256}).

\begin{figure*}[ht]
  \centering
  \includegraphics[width=0.49\linewidth]{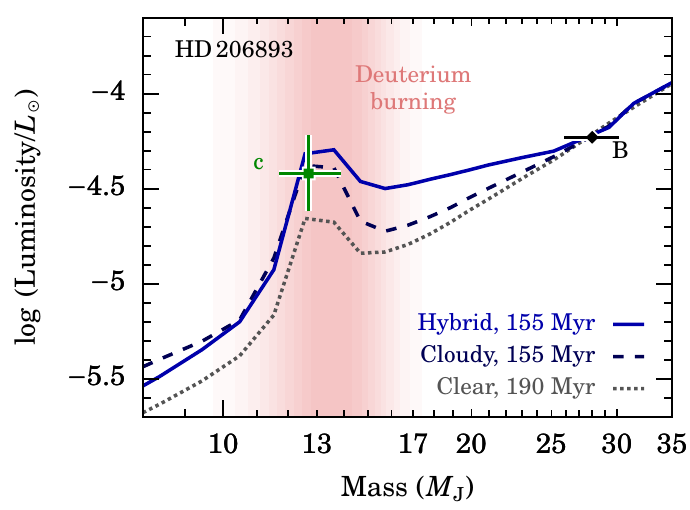}
  \includegraphics[width=0.49\linewidth]{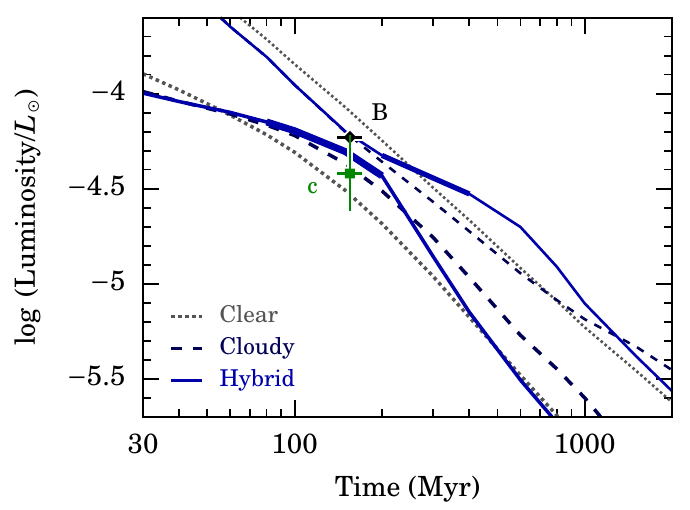}
  \caption{\textit{Left:} Isochrones from \citet{saumonEvolutionDwarfsColorMagnitude2008} matching the dynamical mass and bolometric luminosity of HD\,206893B, with the respective age indicated in the legend.
  The deuterium-burning mass limit, which depends on bulk planetary properties, is shown as a fuzzy pink region \citep{spiegel2011}. 
  \textit{Right:} Corresponding cooling curves for the best-fit masses. The system age of $155\pm15$~Myr comes from our isochrone analysis (see text).
  The thick part of the hybrid cooling curves highlights $1400~\mathrm{K}>\Teff>1200~\mathrm{K}$, with cloudy (clear) atmospheres at earlier (later) times.
  }
  \label{fig:isocc}
  \end{figure*}

In Figure~\ref{fig:isocc} (left panel), we show isochrones for the three varieties of \SMref models: clear (190~Myr), cloudy (155~Myr), and hybrid (155~Myr). They all pass through the well-constrained luminosity of B at its dynamical mass, allowing for relatively tight constraints on the age of the system.
A more careful statistical assessment is planned for future work  
but isochrones at $155\pm15$~Myr ($155\pm20$~Myr) for hybrid (cloudy) models pass within one mass errorbar of HD\,206893B.
Figure~\ref{fig:isocc} shows the important finding that both the cloudy and the hybrid models are able to fit both objects, but that the clear models of \SMref are somewhat inconsistent (to about one sigma)  
with the luminosity of c. This applies also to the COND03 models, which would require an age of 195~Myr as we verified separately. The 140--170 Myr isochrones that pass through one mass standard deviation of HD\,206893B are also easily consistent with the luminosity of HD\,206893c (not shown).


Figure~\ref{fig:isocc} also displays cooling curves for the three \SMref-family models, using the best-fit masses. 
When using clear atmospheres, the cooling curves of HD\,206893c are somewhat underluminous, but are overluminous for HD\,206893B.  
Deuterium burning barely plays a role near the current age of the higher-mass object, the nuclear fuel being instead spent at an earlier age $\sim10$~Myr (e.g., \citealp{burrows01}).
Cloudy atmospheres, whether assumed to be present at all times or only down to $\Teff=1400$~K (hybrid sequence), change the relative cooling so as to ``pinch'' the cooling curves together.%

Interestingly, according to the hybrid model of \SMref, HD\,206893c might be in the process of losing its clouds while its luminosity would be the same as an object that is still cloudy. Also for HD\,206893B, the transition from a cloudy to a clear atmosphere would change significantly the luminosity compared to a pure cloudy cooling sequence only at later ages $t\gtrsim500$\,Myr. 



\newcommand{\BigCommentingOut}[1]{}


\BigCommentingOut{


\subsection{Models Incorporating Enhanced Helium Abundance}

\textcolor{red}{\textbf{[OLD SECTION; will be in part recycled, in part removed]}}

Despite the significantly different dynamical masses derived for HD\,206893B and c, the relative similarity in bolometric luminosities for both components is striking: $\log(L/L_\odot)\simeq-4.23$ versus $-4.42$ for B and c, respectively.  Thus, assuming a coeval formation for the two objects, standard hot-start evolutionary models \citep[e.g.,][]{bmh97, baraffeEvolutionaryModelsCool2003,bha15} cannot simultaneously provide a solution for the brightness of both objects.  Adjusting the initial entropy \citep{sb12, mc14} of either object within these evolutionary models does not ease the tension.  Indeed, an elevated initial entropy will not increase the luminosity of c at its present age of $\approx$200\,Myr because this age is already greater than its initial cooling timescale \citep{mc14}. Similarly, no reasonable lower initial entropy for B could sufficiently delay the begin of deuterium burning 
so that it would be observed in the rising part of a ``deuterium flash'' (see Fig.~8 of \citealt{mc14} or Fig.~13 of \citealt{bcm14}).

\noindent
\textcolor{green}{TODO: add sentences about importance of clouds, success of SM08 with other objects? But maybe careful since spectrum here will not match.}

\textcolor{blue}{\textbf{Instead, we turn towards the models of \citet[][hereafter \SMref]
{saumonEvolutionDwarfsColorMagnitude2008}. \SMref provide cooling sequences for objects with a cloudy or a clear (cloud-free) atmosphere at all ages, which they show is very similar to respectively AMES-COND \citep{baraffeEvolutionaryModelsCool2003} or DUSTY00 \citep{chabrierEvolutionaryModelsVery2000}. But the particular interest of \SMref is that they also calculate a ``hybrid sequence'' using a transition from cloudy at high effective temperature $\Teff\geqslant1400$~K to cloud-free at lower $\Teff\leqslant1200$~K. This has the potential of delaying the cooling differently for both objects and could provide an explanation of our observations. These models have been already applied to several objects (e.g., \citealp{dupuy22_1256}).}}

[Interesting but old:]\\
\textcolor{red}{Instead, the puzzle of a single set of models being unable to simultaneously account for the luminosities of both objects can be addressed by letting c burn deuterium significantly. With a classical value of $(13\pm0.8)~\MJ$ for the mass above which an object burns more than 50\%\ of its initial deuterium content \citep{sbm11}, nuclear reactions do not slow down the cooling of c significantly even at an upper-one-sigma mass of 13.4~\MJ. However, \citet{sbm11} and \citet{mm12} have shown that varying the helium, deuterium, or metal content of an object can shift the D-burning mass boundary significantly, with 11.8~\MJ a plausible value.}
%
If both objects are co-eval, the proximity of c to the deuterium-burning limit makes it possible to let it be in a deuterium-burning phase by varying model parameters while not affecting the modeled luminosity of B. Because of its higher mass, it is already past its deuterium-burning phase, which, for hot starts, occurs earlier at high mass.

\textcolor{red}{
The simplest adjustment is to increase the bulk deuterium abundance but quantitatively this is problematic. 
From our tests (not shown), a model that is consistent with the brightness of HD\,206893c would call for a deuterium number fraction relative to hydrogen several times higher than in the local interstellar medium, (D/H)$_{\rm ISM}$ = (2.0$\pm$0.1)$\times$10$^{-5}$ \citep{psf10}, or the (D/H)$_{\rm BBN}$ = (2.57$\pm$0.03)$\times$10$^{-5}$ value from Big Bang Nucleosynthesis (BBN) estimates \citep{fms14}. The latter is effectively an absolute upper limit because no process can increase the bulk abundance of deuterium; it can only be destroyed.}

\textcolor{red}{
A more realistic route is to enhance the helium mass fraction, $Y$, to explain the overluminosity of HD\,206893c. An elevated helium fraction can serve as a proxy for an overall heavier composition \citep{bacb06,bcb08}, representing an approximately normal helium abundance along with a higher bulk metallicity. Physically, a heavier composition leads to a more compact planet with correspondingly higher internal temperatures, in turn leading to enhanced deuterium burning at lower masses. 
Also, deriving a heavy element content (helium and metals) abundance from such models could help place constraints on formation locations within the primordial protoplanetary disk \citep{cpa16, mmb22}.} 

\textcolor{red}{
Therefore, we use the models of \citet{mc14} to model the luminosity of substellar objects as a function of their mass and age. 
We have extended them to use the \citet{baraffeEvolutionaryModelsCool2003} atmospheric boundary condition as in the BEX models \citep{marleauExploringFormationCore2019},  
and have generated grids of models with different $Y$ values and going up to high masses.
Figure~\ref{Dburning} shows an example of these model isochrones at an age of 170\,Myr for hot starts (here, an initial entropy of 13.5\,\Sunits) and a helium fraction ranging from $Y$=0.25 to 0.40. For reference, a curve is also shown assuming a younger age of 50\,Myr, consistent with the luminosity of the c companion but not B. It is notable that the curve assuming a 170\,Myr age and a baseline abundance of $Y=0.25$ cannot simultaneously account for the luminosities of HD\,206893B and c (at the level of $\sim$2$\sigma$ in the mass coordinate).  However, by varying the level of helium enrichment, solutions can be found assuming an age of 170\,Myr for both. This is because the luminosities of lower-mass objects ($\sim$12--17\,\MJ) near the deuterium burning limit can vary at a fixed age of 170\,Myr by a factor of a few up to an order of magnitude depending on the helium fraction.  Thus, using these models, we find that a 170\,Myr model assuming a helium fraction of $Y=0.33$ could
account for the luminosities of both components of the HD\,206893 system. For this exploratory analysis we match the data only at approximately 1\,$\sigma$ of both mass values, and will attempt a more careful fit in a future work. 
}

\textcolor{red}{
As a check, we have run similar models using the MESA stellar evolution code \citep{paxton11,paxton19}. The results are largely in line with the discussion above. Indeed, adjusting these parameters within MESA (e.g., D/H, helium abundance, etc.) one by one, we can reproduce the luminosity for HD\,206893c to within a factor of $\sim$2. These tests give confidence that our results are not representing any approximations inherent in the \citet{mc14} models. }

Importantly, since the luminosity of HD\,206893B is not strongly affected by the initial deuterium or the helium abundance in the models, this object can be considered as an independent ``chronometer'' of the system. Assuming the two members of the system to be coeval\footnote{Even for an age of 50\,Myr, the formation time of at most several million years is a negligibly short fraction of the system's age.}, this lets our models set constraints on the bulk properties of HD\,206893c. The pink shaded region in Fig.~\ref{Dburning} highlights approximately the mass range in which deuterium burning keeps the luminosity significantly higher at a given age than for neighbouring masses.
It is a clear prediction of these models that after several hundred million years the deuterium burning will cease, leaving HD\,206893c (again, in this plot at the upper 1\,$\sigma$ limit of the mass) squarely in the
luminosity regime typically associated with planets rather than brown dwarfs.  

\textcolor{red}{
It is not clear what process(es) could increase the bulk concentration of helium to $Y=0.33$, but our result should be taken more broadly to indicate that non-standard parameter values for the cooling curves are required to reproduce the observations. In a forthcoming paper, this will be given a detailed statistical treatment by varying systematically the initial entropies of the objects, their core masses, deuterium abundances, and so on. In any case, the distance in Fig.~\ref{Dburning} between the data points and the standard models is too large to be explained by simple measurement errors or small model uncertainties.  
}

\begin{figure*}[ht]%
\centering
\includegraphics[width=12cm]{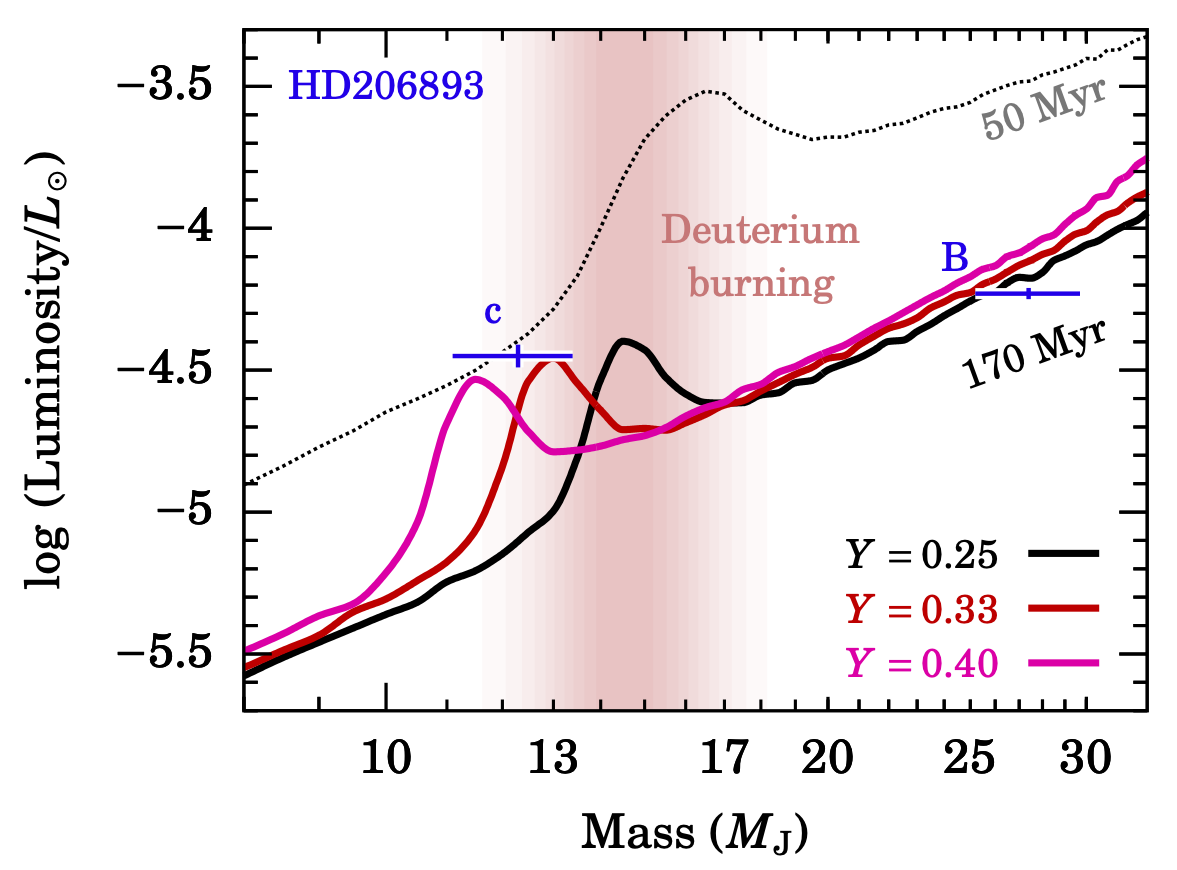}
\caption{
Hot-start isochrones from \citet{mc14} but coupled to \citet{baraffeEvolutionaryModelsCool2003} as in \citet{marleauExploringFormationCore2019} with helium fraction ranging from $Y=0.25$ to 0.40 (solid lines), compared to the dynamical masses and retrieved luminosities of HD\,206893B and c. With the baseline helium content $Y=0.25$, the 170\,Myr isochrone (solid black curve) matches the bolometric luminosity of HD\,206893B assuming a mass of 27\,\MJ but does not match c, and conversely for the 50\,Myr isochrone (dotted black, with $Y=0.25$). Increasing the helium abundance to $Y=0.33$ could  nearly explain the measurements of both components of the HD\,206893 system.
}
\label{Dburning}
\end{figure*}

}   

\begin{table}[]
    \centering
    \begin{tabular}{lcc}
\hline
&{\textbf{HD\,206893c} } & {\textbf{HD\,206893B}}\\ 
\hline\hline
\vspace{0.2cm}
   $M$ [\MJ]           & 12.7$^{+1.2}_{-1.0}$ (11.5$^{+2.4}_{-2.2}$)     &        28.0$^{+2.2}_{-2.1}$ (26.2$^{+3.7}_{-3.6}$)   \\
   \vspace{0.2cm}
   $a$ [au]            & 3.53$^{+0.08}_{-0.06}$ (3.68$^{+0.12}_{-0.09}$) &       9.6$^{+0.4}_{-0.3}$ (9.7$^{+0.8}_{-0.4}$)                 \\
   \vspace{0.2cm}
   $e$            & 0.41$^{+0.03}_{-0.03}$ (0.36$^{+0.05}_{-0.06}$) &        0.14$^{+0.05}_{-0.05}$ (0.13$^{+0.06}_{-0.09}$)                 \\
   \vspace{0.2cm}
   $R$ [\RJ] &         1.46$^{+0.18}_{-0.06}$   &  1.25$^{+0.02}_{-0.02}$\\
   \vspace{0.2cm}
   $T_{\rm eff}$ [K]     &  $1181.9^{+20.6}_{-53.9}$ &  $1429.2^{+5.6}_{-6.2}$  \\ \vspace{0.2cm}
   $\log g$           &  $4.08^{+0.10}_{-0.18}$   &  4.66$^{+0.04}_{-0.04}$ \\ \vspace{0.2cm}
   $\log(L/L_\odot)$     & $-4.42^{+0.02}_{-0.01}$  & $-4.23^{+0.01}_{-0.01}$  \\               
   $\textrm{[Fe/H]}$     & $0.28^{+0.02}_{-0.04}$   & 0.27$^{+0.02}_{-0.05}$    \\ \vspace{0.1cm}
   Contrast     & $8.2\times10^{-5}$     &  $1.6\times10^{-4}$ \\
\hline
\hline
    \end{tabular}
    \caption{Orbital and physical parameters for HD\,206893c, and re-evaluated parameters of HD\,206893B. $M$, $a$ and $e$ are obtained from the dynamical fits. Values not listed in parentheses are those which use both the astrometry and RVs during the dynamical orbit fit (and are the most trustworthy), while those in parentheses exclude the RV measurements. $R$, \Teff, $\log g$ and $\log(L/L_\odot)$ are obtained from fitting the spectrophotometry. The fits use a grid of DRIFT-PHOENIX models assuming no additional dust, and using mass priors of $26\pm2$ and $12\pm1$\,\MJ for B and c, respectively. The contrast refers to that measured in the K-band.  The luminosity uncertainties
    for HD\,206893c are only the statistical ones
    within the fits to the DRIFT-PHOENIX models.
    Further analyses should use instead the more appropriate value of 0.2~dex
    %
    that reflects the systematic modelling uncertainties (Section~\ref{sec:spectra_modelling} and \ref{sec:disc}).
    }
    \label{tab:params}
\end{table}

\section{Discussion}
 \label{sec:disc}
The highly precise astrometry delivered by GRAVITY ($\sim$50--100\,$\upmu$arcsec)
let us derive
dynamical masses for the objects in the HD\,206893 system that are precise at the level of 5--10\%.  Continued astrometric monitoring of both objects could, of course, lead to tighter constraints on the derived dynamical masses, and thus yield a more precise age determination. 
In combination with \SMref models, the masses and luminosities reveal that both cloudy models and hybrid cloudy models featuring a cloudy-to-clear transition describe well the cooling of both objects, and that a clear atmosphere for both objects is not consistent with the observations.

Thus already the bolometric luminosities, independently of spectroscopy, disfavour the scenario in which both companions have cooled with a clear atmosphere up to now.  However, while a 140-Myr \SMref clear isochrone is overluminous at the mass of of HD\,206893B, it does match the luminosity of HD\,206893c. Thus both objects could have different degrees of cloudiness.

%
It is unlikely that unknown systematic errors in our dynamical analysis are leading to calculated values of the mass of HD\,206893B and c that are erroneous.  Our calculation of the dynamical mass uses a combination of stellar astrometry that may have systematic uncertainties that are not well understood, as well as RV measurements that are impacted by stellar activity \citep[e.g.,][]{glb19}. However, Table~\ref{tab:params} shows that excluding the RV data from our fits gives very similar calculated values for the mass of HD\,206893c  (12.7$^{+1.2}_{-1.0}$\,\MJ versus 11.5$^{+2.4}_{-2.2}$\,\MJ with the RVs excluded).  It should be noted that in this analysis we did not model the stellar activity in our orbit fit, which may contribute to a slight bias.  However, it is unlikely that such unknown systematic errors within our analysis have significantly biased our calculated masses, especially given that our measured dynamical mass of HD\,206893c is very much consistent with the predictions presented in \citet{glb19} and \citet{kls21}. At most, these systematics could have led to underestimated astrometric uncertainties, but this would not have changed the overall conclusions of this study.  


As discussed previously, we compared the spectra from the different atmosphere model families, including \SMref (spectra courtesy of D.\ Saumon, 2022), to the GRAVITY data. In several cases the shape is reproduced only approximately, and the shape of the \SMref models is not red enough. However, there is a large amount of correlated noise within the GRAVITY spectrum, precluding a simple statistical analysis. To address this, taking a conservative approach to estimate the systematic modelling errorbar on the luminosity of HD\,206893c, we have therefore adopted the scatter in the \Lbol value from an ensemble of different models.
We allow the $K$-band shape not to match exactly as a way for accounting for detailed physics currently missing in the atmospheric models, while assuming that they capture the overall luminosity evolution. The spectral energy distribution over a small wavelength range, for instance the $K$ band, is much more model-dependent than the overall bolometric flux.

Finally, in Appendix~\ref{sec:dynamics} we describe our  study of the dynamical stability of the HD\,206893 system.  Our analysis demonstrates that while the architecture of this system may resemble that of the $\beta$\,Pictoris system, the HD\,206893 system is much closer to instability.

\section{Conclusions}
We have presented the discovery of an inner ${\sim}12-13$\,\MJ exoplanet at 3.5\,au in the HD\,206893 system, making this one of the first directly imaged ``hybrid'' planetary systems containing both a brown dwarf \citep{mhc17} and a bona fide exoplanet.  This is also an example of a discovery of an additional companion following a previous discovery of an outer one \citep[e.g.,][]{lmr19, nll20}. The highly precise astrometry delivered by GRAVITY ($\sim$50--100\,$\upmu$arcsec) has enabled us to derive precise dynamical masses of 12.7$^{+1.2}_{-1.0}$\,\MJ for HD\,206893c and 28.0$^{+2.2}_{-2.1}$\,\MJ for HD\,206893B. The dynamical mass of B, combined with its luminosity, allows us to immediately and unambiguously establish a robust age of $\sim$150\,Myr for this system. 
Further, having both an EGP and a BD in the same system, with a common age, HD\,206893\, will serve as an extraordinarily valuable system for the purpose of furthering our understanding of planet formation, and highlighting distinctions between the formation pathways of EGPs and BDs.   

%

We have used the theoretical atmospheric models from \SMref, which self-consistently treat deuterium burning in substellar objects to show the bolometric luminosities of both objects can be well-modelled by cloudy/hybrid models that allow us to fine tune the age value to 155$\pm$15\,Myr. 
Besides residing in a hybrid exoplanetary system, HD\,206893c is an object straddling the deuterium-burning limit, and thus is an ideal laboratory to establish the precise mass where this process can occur. 
In addition to having a dramatic impact on its evolution, the duration and extent of deuterium burning in a substellar object may give valuable clues to its initial conditions and internal structure \citep[e.g., the presence/lack of a rocky core, initial entropy, and initial deuterium abundance,][]{sb12,mm12,bdl13, mordasini13}.  Ours is the first such study to begin addressing these questions, and additional studies like this may help to craft a more robust discrimination between EGPs and BDs.

Finally, in addition to being only the second directly imaged exoplanet whose presence was hinted at using the RV method, our precise dynamical mass for HD\,206893c means that this effort potentially represents the first discovery via direct imaging of a bona fide exoplanet that partially relies on precise measurements of the host star astrometry from the \textit{Gaia} mission.  This complements numerous other recent discoveries of substellar companions (e.g., \citealp{bonavita22, kct22, franson22, fbb22}; Currie et al.\ in prep.) whose discovery also used \textit{Gaia} astrometry.  Using precise \textit{Gaia} astrometry of the host star to point the way to orbiting planets suitable for direct imaging is expected to be one of the primary strategies for direct exoplanet detection and characterization going forward, thereby ending the current era of ``blind'' direct imaging EGP searches that have had notoriously low detection rates \citep[e.g.,][]{bn18}.  
Secondly, recent exoplanet direct imaging surveys have had limited sensitivity \citep[e.g.][]{ndm19, vfm21} to the peak of the orbital distribution of giant exoplanets that coincides with the location of the water ice line at 2--3\,au for solar-type stars \citep[e.g.,][]{fmp19,fjm19,frh21}. However, upcoming interferometric observations using \textit{JWST} \citep[][]{rhs22, hcr22} are likely to have the combination of sufficient sensitivity and resolution to reach these orbital zones, providing complementary characterization at 3-5\,$\mu$m.  Until then, along with the discovery of $\beta$\,Pic\,c \citep[][]{lmr19,nll20} this discovery of an exoplanet at 3.5\,au is more evidence that optical interferometry now enables direct characterization of these planets at the ice-line orbital separations of 2--3\,au \textit{where they form.}

\begin{acknowledgements}
We are very grateful to D.\ Saumon for helpful and rapid replies, and kindly sharing data.
G-DM acknowledges the support of the DFG priority program SPP 1992 ``Exploring the Diversity of Extrasolar Planets'' (MA~9185/1) and from the Swiss National Science Foundation under grant
200021\_204847 ``PlanetsInTime''.
Parts of this work have been carried out within the framework of the NCCR PlanetS supported by the Swiss National Science Foundation.
      P.M.\ acknowledges support from the European Research Council under the European Union's Horizon 2020 research and innovation program under grant agreement No.~832428-Origins.
S. A. acknowledges support from the European Research Council under the European Union’s Horizon 2020 research and innovation program under grant agreement No. 865624-GPRV.
NZ acknowledges support from the UK Science and Technology Facilities Council (STFC) under Grant Code ST/N504233/1, studentship no. 194772.
This research has made use of the SIMBAD database, operated at CDS, Strasbourg, France.
This work has made use of data from the European Space Agency (ESA) mission {\it Gaia} (\url{https://www.cosmos.esa.int/gaia}), processed by the {\it Gaia} Data Processing and Analysis Consortium (DPAC,\url{https://www.cosmos.esa.int/web/gaia/dpac/consortium}). Funding for the DPAC has been provided by national institutions, in particular the institutions participating in the {\it Gaia} Multilateral Agreement.

\end{acknowledgements}

\bibliographystyle{aa}
\bibliography{MasterBiblio_Sasha.bib,tomas_ref,SylvestreLibrary}




   
  



\begin{appendix} 

\section{Detailed orbit fitting results}\label{sec:orbit_fit_detail}
We use the default orbital basis in \citet{bwa20}. The only difference is that $\tau$, the epoch of periastron after a given reference epoch in units of orbital period, uses a reference epoch of MJD 59000. To fit the radial velocity data, we also include two free parameters to fit for a bulk offset in the radial velocity ($\gamma_{RV}$) and a jitter term to inflate the error bars on the radial velocity data to better account for stellar activity ($\sigma_{RV}$). We list the median and 68-percentile credible intervals centered about the median as the error bars in Table \ref{table:detailed_orbit}. We also computed orbital periods ($P_B$ and $P_c$) for both companions as well as the mutual inclination of their orbits and listed those in Table \ref{table:detailed_orbit}. 

\begin{table}[]
    \centering
    \begin{tabular}{ccc}
    \hline
 Quantity  & Prior & Posterior \\
    \hline
    \hline
 $a_B$ (au)               & LogUniform(1, 100) & $9.6^{+0.4}_{-0.3}$  \\
 $e_B$                    & Uniform(0, 1) & $0.14^{+0.05}_{-0.05}$  \\
 $i_B$ (\degr)            & $\sin(i)$ & $146.8^{+3.3}_{-3.9}$  \\
 $\omega_B$ (\degr)       & Uniform(0, 2$\pi$) & $183^{+10}_{-10}$ \\
 $\Omega_B$ (\degr)       & Uniform(0, 2$\pi$) & $73.5^{+5.0}_{-3.4}$ \\
 $\tau_B$                 & Uniform(0, 1) & $0.308^{+0.030}_{-0.020}$ \\
 \hline
 $a_c$ (au)               & LogUniform(1, 100) & $3.53^{+0.08}_{-0.06}$ \\
 $e_c$                    & Uniform(0, 1) & $0.41^{+0.03}_{-0.03}$ \\
 $i_c$ (\degr)            & $\sin(i)$ & $150.9^{+2.9}_{-3.0}$ \\
 $\omega_c$ (\degr)       & Uniform(0, 2$\pi$) & $46^{+8}_{-8}$ \\
 $\Omega_c$ (\degr)       & Uniform(0, 2$\pi$) & $89.1^{+6.9}_{-7.0}$ \\
 $\tau_c$                 & Uniform(0, 1) & $0.687^{+0.012}_{-0.011}$ \\
 \hline
 Parallax (mas)         & $\mathcal{N}$(24.5275, 0.0354) & $24.5276^{+0.0358}_{-0.0367}$ \\
 $\gamma_{RV}$ (m/s)         & Uniform($-5000$, 5000) & $140^{+11}_{-13}$ \\
 $\sigma_{RV}$ (m/s)         & LogUniform(0.1, 100) & $38^{+3}_{-3}$ \\
 $M_{B}$ (\MJ)  & Uniform(1, 50) & $28.0^{+2.2}_{-2.1}$ \\
 $M_{c}$ (\MJ)  & Uniform(1, 50) & $12.7^{+1.2}_{-1.0}$ \\
 $M_{*}$ ($M_\odot$)  & $\mathcal{N}$(1.29, 0.10) & $1.32^{+0.07}_{-0.05}$ \\
 \hline
 \multicolumn{3}{c}{Derived Parameters} \\
 \hline
 $P_B$ (yr)         & -- & $25.6^{+1.2}_{-1.2}$ \\ 
 $P_c$ (yr)         & -- & $5.74^{+0.12}_{-0.10}$ \\ 
 Mutual Inc.\ (\degr)    & -- & $9.1^{+5.6}_{-4.9}$ \\ 
 \hline
 \end{tabular}
 \caption{Full set of fitted orbital parameters for HD\,206893B and c, in Jacobi coordinates, with some derived parameters included.}
 \label{table:detailed_orbit}
\end{table}

\section[HD 206893 Debris Structure]{HD\,206893 debris structure}
HD\,206893 also hosts a prominent debris disk indicated by a high fractional luminosity of $L_{\rm dust}/L_{\star}$=2.3$\times$10$^{-4}$ \citep{mad06}.  The disk was originally identified by the \textit{ISO} instrument \citep{s00}, and subsequently characterised in greater detail with \textit{Spitzer} \citep{cmk14}, \textit{Herschel} and \textit{ALMA} observations \citep[][]{mhc17,mzf20, nhf21}. 
\citet{mzf20} imaged the disk with ALMA, finding that the radial structure is broad, extending from 30 to 180\,au, with a 27\,au wide gap at 74\,au.
We have re-analyzed the \textit{Spitzer} IRS spectrum of HD\,206893 to recalculate the temperature of the 499\,K warm dust component found by \citet{cmk14} to be consistent with the methods described in \citet{kw14}. 
Our reconstruction of the star + disk spectrum using the methods described by \citet{yks19} is consistent with a single dust component with a temperature of approximately 50\,K. The inner disk edge is therefore likely consistent with inner truncation through interaction with HD\,206893B.
Following the method presented in \citet{ldm18}, we analytically estimated that the region cleared from dust due to the presence of the two planets extends from $\sim$1.5 au to $\sim$15.5 au. Since the clearing zone of the planets cannot cover the outer extent of the gap, detected at 30 au \citep{mzf20}, this could imply the presence of a third smaller planet responsible for shaping the inner edge of the disk


\section{Dynamical analysis}\label{sec:dynamics}
\begin{figure}[ht]
	\centering
	\includegraphics[width=0.8\linewidth]{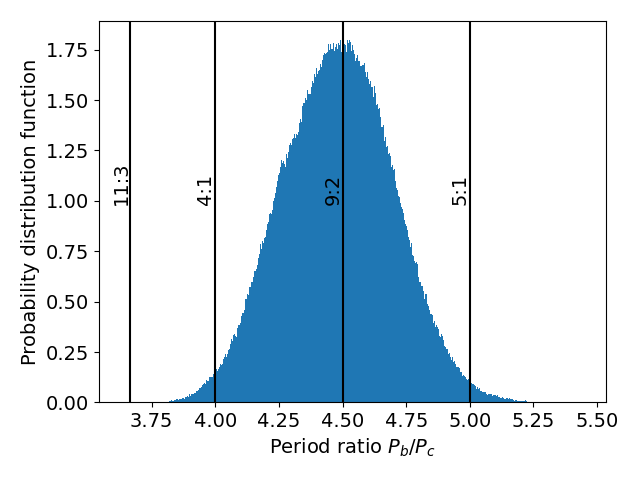}
	\caption{Probability distribution of HD\,206893B and c period ratios from the orbital fit.}\label{fig:periodratio}
\end{figure}

Various dynamical considerations can be made from the results of the orbital fit. First, we note that the orbital architecture of the companions does not exhibit a clear resonant configuration. Indeed, as can be seen in Fig.~\ref{fig:periodratio}, the probability distribution of the period ratio lies between the 4:1 and 5:1, and favours instead 9:2, which corresponds to a $7^\mathrm{th}$ order mean-motion resonance (MMR). Such a high-order MMR is not strong enough to induce the libration of a resonant angle, especially when the companions are so massive. Thus, we do not expect resonant capture or phase protection to play a role in the shaping or the stability of the companions orbits.
	
\begin{figure*}[ht]
		\centering
		\includegraphics[width=0.45\linewidth]{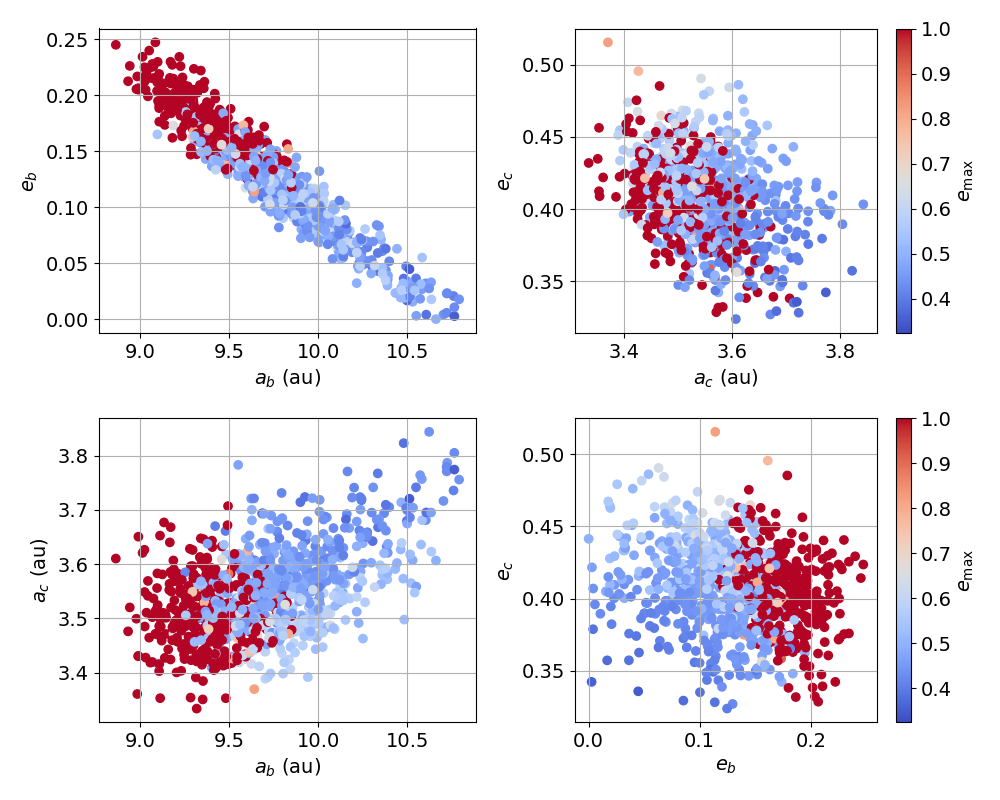}
		\includegraphics[width=0.45\linewidth]{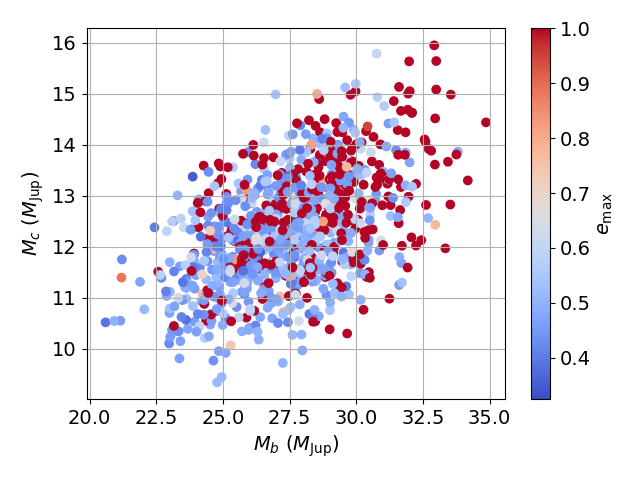}
		\caption{Maximum eccentricity reached by either companion in $N$-body simulations with initial conditions representative of the orbital solutions. The simulations are performed for $10^5$ yr using the Mercurius integrator.}\label{fig:maxe}
	\end{figure*}

If the orbital configuration looks very similar at first to the $\beta$ Pictoris system, the dynamics of the HD\,206893 system is much closer to instability. This is due to the smaller separation between the two companions, and the higher mass and eccentricity of the inner companion. We sample randomly $1000$ orbital solutions from the orbital fit, and perform $N$-body simulations starting from these solutions and lasting for $10^5$\,yr. 
The simulations are performed using the Mercurius integrator in the Rebound package \citep{reinREBOUNDOpensourceMultipurpose2012,reinHybridSymplecticIntegrators2019}, with a time-step of $0.1$\,yr. We monitor the eccentricities of both companions and stop the simulations if one of them reaches one. 
The results are displayed on Fig.~\ref{fig:maxe}: the simulations exclude $e_B > 0.15$ and $a_B < 9.5$\,au, and they disfavor the highest masses. 
Extending the integration to 10$^6$\,yr does not significantly change the results, so that we can assume the stable configurations remain stable for longer than the integration time. The probability that we are observing the system within 10$^5$\,yr of a companion's ejection is very unlikely given its age, therefore the system is likely long-term stable.
Despite the relatively large mutual separation, the system can be considered compact, as the orbits of the companions are close to instability. It suggests that ejection may be a likely outcome of the formation of two massive companions in planetary-like orbits.

Another important characteristic of this hierarchical three-body system is its relative coplanarity. Table~\ref{table:detailed_orbit} shows indeed that the mutual inclination between both orbital planes is $\sim$15$\degr$ at most. This actually reinforces the similarity with the $\beta\:$Pictoris system. This is also low enough to keep this system far enough from Kozai--Lidov resonance \citep{k62}. That dynamical mechanism typically concerns nested orbits of different sizes having a sufficient enough mutual inclination ($\ga 39\degr$ theoretically). When this occurs, the inner orbit is subject to large amplitude eccentricity oscillations that can drive it to very high values close to unity. This can lead to direct instability or to a potential collision with the central star due to a decaying periastron. The latter fate can be prevented by tides within the inner orbit, but even in that case, the orbit does not remain as it is and furthermore shrinks significantly \citep{ft07,bbm12}. The mutual inclination determination (Table~\ref{table:detailed_orbit}) shows that there is no risk of Kozai--Lidov resonance here, which reinforces the likely stability of the system. This also favours the stability of the debris disk, as disk particles could be affected by Kozai--Lidov mechanism as well.

\begin{figure}[ht]
		\centering
		\includegraphics[width=\linewidth]{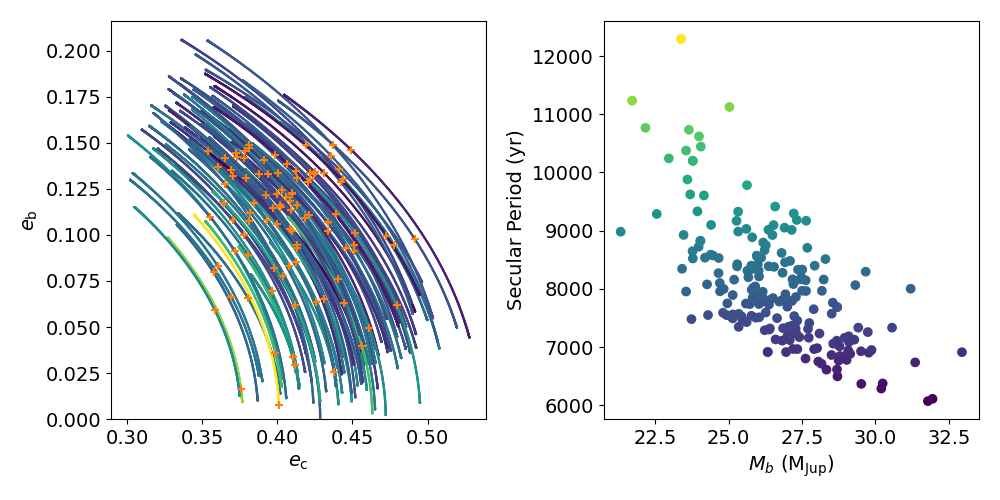}
		\caption{Secular evolution of the eccentricities of $B$ and $c$ due to their mutual interaction. Left is the trajectory in $(e_B, e_c)$ phase space, starting from the observed values (orange crosses, which are random solutions to the orbital fit using all available data). We exclude the initial conditions with $e_B > 0.15$ and $a_B < 9.5$\,au. The colours represent the corresponding  eccentricity evolution period (secular period). The right panel represents the aforementioned period with respect to the mass of b.}\label{fig:secular}
	\end{figure}

	Finally, we note that the stable configurations are just snapshots of a system with complex and evolving dynamics. We use the approach described in \citet{lwr21} to estimate the secular eccentricity variation and the associated period. We exclude solutions with $e_B > 0.15$ and $a_b < 9.5$ au to ensure stability. The results are displayed on Fig.~\ref{fig:secular}. The eccentricity of HD\,206893B can periodically vanish, while $e_c$ can reach values as high as $0.5$. These variations are expected to occur over timescales of ${\sim} 10^4$\,yr. This is important to keep in mind when discussing the formation of the orbital architecture (as it could have formed at any point of the secular cycle), or the impact of the companions on the other components of the system such as a debris disk or additional planets.

\end{appendix}

\end{document}